\documentclass[smallcondensed,envcountsect,envcountsame,envcountreset]{svjour3} %[release-date]
\usepackage[margin=1in]{geometry}
\usepackage{amssymb}
\usepackage[T1]{fontenc}

\usepackage{hyperref}

\def\Eg{E{.}g{.}}
\def\eg{e{.}g{.}}
\long\def\comment#1{}
\def\widetilde{\mathaccent"0365 }

\catcode`@=11
\def\eqalign#1{\null\,\vcenter{\openup1\jot \m@th
   \ialign{\strut \hfil$\displaystyle{##}$ & $\displaystyle{{}##}$\hfil
      \crcr#1\crcr}}\,}
\DeclareRobustCommand{\twodots}{\t@urdots}
\def\t@urdots{\mbox{\kern1\p@\vbox to 1ex{\hbox{.}\vss\vss\hbox{.}}}}
\DeclareRobustCommand{\treedots}{\th@urdots}
\def\th@urdots{\mbox{\kern1\p@\vbox to 1.3ex{\hbox{.}\vss \hbox{.}\vss\hbox{.}}}}
\catcode`@=12

\def\otd{\mathop{\;\raise-1.65pt\hbox{\Large$0$}\mkern-10.2mu \treedots}\;}

\def\eref#1{(\ref{#1})}
\def\eqdef{{\;\mathop{:=}\;}}

\def\dd{(\cdot,\cdot)}
\def\demi{{1\over 2}}
\let\ds\displaystyle
\def\dvg{{\rm div}}
\def\Im{{\rm Im}}
\def\NN{\mathbb{N}}
\def\RR{\mathbb{R}}
\def\RRn{{\RR^n}}
\def\RRnt{{\RR^n_t}}
\def\RRntz{{\RR^n_\tz}}
\def\RRns{{\RR^{n*}}}
\let\pa\partial
\let\ora\overrightarrow

\def\qavec{\quad\hbox{with}\quad}
\def\qand{\quad\hbox{and}\quad}
\def\qif{\quad\hbox{if}\quad}
\def\qthen{\quad\hbox{then}\quad}
\def\qwhere{\quad\hbox{where}\quad}

\def\tz{{t_0}}
\def\tu{{t_1}}
\def\td{{t_2}}
\def\bigC{\hbox{\large$\cal C$}}
\def\eul{{\cal E}}
\def\Obj{{\it O\!b\!j}}
\def\Omegat{{\Omega_t}}
\def\Omegatau{{\Omega_\tau}}
\def\Omegatz{{\Omega_\tz}}
\def\Phitz{{\Phi^\tz}}
\def\Pobj{{P_\Obj}}

\def\OmegaE{{\Omega_\calE}}

\def\rar{\rightarrow}
\def\vgamma{{\vec\gamma}}

\def\Tuuot{{T^1_1(\Omegat)}}
\def\Tzdot{{T^0_2(\Omegat)}}
\def\Tdzot{{T^2_0(\Omegat)}}

\def\Phitz{{\Phi^\tz}}
\def\Phitzt{{\Phi^\tz_t}}

\def\tPhi{{\widetilde\Phi}}

\def\Ctzt{{C^\tz_t}}
\def\Etzt{{E^\tz_t}}
\def\Fttau{{F^t_\tau}}
\def\Ftz{{F^\tz}}
\def\Ftzt{{F^\tz_t}}
\def\Htz{{H^\tz}}
\def\Htzt{{H^\tz_t}}
\def\Stzt{{S^\tz_t}}
\def\Tr{{\rm Tr}}

\def\calL{{\cal L}}
\def\calP{{\cal P}}
\def\calPa{{{\cal P}_a}}
\def\calPe{{{\cal P}_e}}
\def\calPi{{{\cal P}_i}}
\def\calPpres{{{\cal P}_{\it pres}}}
\def\calV{{\cal V}}

\def\pint{{p_{\rm int}}}

\def\odd{\mathop{\;\raise-1.55pt\hbox{\large$0$}\mkern-9.5mu \twodots\;}}

\def\sumin{{\sum_{i=1}^n}}
\def\sumim{{\sum_{i=1}^m}}
\def\sumjn{{\sum_{j=1}^n}}

\def\sumijn{{\sum_{i,j=1}^n}}
\def\sumiJn{{\sum_{i,J=1}^n}}

\def\sumijkn{{\sum_{i,j,k=1}^n}}

\def\diag{{\rm diag}}
\let\eps\varepsilon

\def\tmu{{\tilde \mu}}
\def\uua{{\underline{\underline{a}}}}

\def\uuatzt{{\uua^\tz_t}}
\def\uub{{\underline{\underline{b}}}}
\def\uubtzt{{\uub^\tz_t}}
\def\uueps{{\underline{\underline{\eps}}}}
\def\uukappa{{\underline{\underline{\kappa}}}}
\def\uukappam{{\underline{\underline{\kappa_m}}}}
\def\uukappau{{\underline{\underline{\kappa_u}}}}
\def\uukappad{{\underline{\underline{\kappa_d}}}}
\def\uusigma{{\underline{\underline{\sigma}}}}
\def\uusigmam{{\underline{\underline{\sigma_m}}}}
\def\uusigmatzt{{\uusigma^\tz_t}}
\def\uutau{{\underline{\underline{\tau}}}}

\def\va{{\vec a}}
\def\vb{{\vec b}}
\def\ve{{\vec e}}
\def\vE{{\vec E}}
\def\vf{{\vec f}}
\def\vg{{\vec g}}
\def\vn{{\vec n}}
\def\vu{{\vec u}}
\def\vv{{\vec v}}
\def\vVtz{{\vec V^\tz}}
\def\vVtzt{{\vec V^\tz_t}}
\def\vw{{\vec w}}
\def\vW{{\vec W}}

\title{Virtual Power Principle: A Lie Covariant Approach.
Applications to Non-Linear Elasticity, Turbulence, Visco-elasticity}
\author{Gilles Leborgne}
\institute{G. Leborgne \at ISIMA, University of Clermont Auvergne, 63178 Aubi\`ere Cedex, France
\\Tel.: +33 (0)4 73 40 50 23,
\\\email{gilles.leborgne@uca.fr}}
\date{\today
}
\titlerunning{VPP: Lie Covariant Approach}

\begin{document}

\maketitle

\begin{abstract}
A covariant formulation of the virtual power principle based on Lie derivatives is proposed.
The Lie covariant approach does not require an inner product and the Cauchy deformation tensor to start,
but, at first order in a Galilean Euclidean setting,
gives the usual linear results classically obtained with the  Cauchy deformation tensor.
The Lie approach may also enable to differentiate a fluid from a solid
from an analytical point of view,
and leads to propose a model for hysteresis.
In the non-linear first order case we get covariant models
for visco-elasticity, non linear fluids and non linear elasticity
which differ from usual models.
In the second order case, enriched modelizations are obtained.
\end{abstract}

%\subclass{74A, 76A.}

%Article Highlights:
%$\bullet$ Covariant virtual power principle with Lie derivatives.
%$\bullet$ Application to continuum mechanics of solids.
%$\bullet$ Application to fluid mechanics.

\tableofcontents

\section{Introduction}
\label{secintro}

The starting point is the classical virtual power principle as stated by Germain~\cite{GermainAPPV1973}.
%(recalled at~\S~\ref{seccvpp}).
This principle is based on the ``frame invariance principle'',
cf. Truesdell and Noll~\cite{TruesdellNoll2004} p.~43,
which is given under the ``isometric objectivity'' hypothesis:
``Change of frame are related by rigid transformations combined with a time-shift'' (p.~41), and
``the units of length and time are kept fixed'' (p.~42).
% The principle of frame invariance only considers symmetric stress tensor, since a velocity field $\vv$ and its differential~$d\vv$ are not objective quantities, but ${d\vv + d\vv^T \over 2}$ is ``isometric objective''.

However Marsden and Hughes~\cite{MarsdenHughesMFE} p.~22
indicate that the considered objectivity should be covariant:
``The use of geometry in attempting
to isolate the basic principles that are covariant---that is, that make intrinsic
tensorial sense independent of a preferred coordinate system---automatically
clears up several basic issues. For example, balance of linear momentum does
note make tensorial sense as it stands. However, one can make covariant sense
out of balance energy principles with no reference to rigid body motions.
That is, the Noll and Green--Rivlin--Naghdi program can be done covariantly,
although this is not obvious given the existing literature.''
%And they do it in~\S~3.3. % of~\cite{MarsdenHughesMFE}.

But unlike Marsden and Hughes who use
a metric, and the Cauchy deformation tensor~$C$ seen as the pull-back of the metric,
in this manuscript there is no use of any metric a priori:
It is the objective covariant derivative of~Lie that is used to start. %see~\eref{eq10}--\eref{eqh18}.
(For the Lie derivative and its interpretation which motivates this manuscript,
see appendix~\S~\ref{secdlfei}; % and~\eref{eqdefLiev};
And for a comparison with the Cauchy deformation tensor, see appendix~\S~\ref{remspqsv}).

The presentation is here limited to the affine space $\RRn$ to simplify the writings.
%but a generalization to differentiable manifolds can be done, the formulation being based on Lie derivatives.
After recalling the notations and the classical setting in \S~\ref{secnot} and~\ref{seccvpp}, the Lie derivative approach is introduced in~\S~\ref{secppvl}.
The first order, linear, is given in~\S~\ref{secpfei1}, and with a Galilean Euclidean setting the classical formulation is recovered.
Applications to Newtonian fluids and elastic solids is proposed in \S~\ref{exafs} and~\S~\ref{secsel0}.
And for solids the Lie approach immediately gives
a linear model for the elastic stress tensor: %see~\eref{eqevs1c} or~\eref{eqev2},
To be compared with the classical approach where the linear model is obtained by
linearization of the (quadratic type) Cauchy deformation tensor $C=F^T.F$, see appendix~\S~\ref{secsel}.
Moreover the Lie approach may give a way to differentiate fluids from solids,
and thus, in the first order case give a hysteresis model, see \S~\ref{secfvs}.
And the Lie approach may also provide a simple characterization of hyper-elasticity, see~\S~\ref{secahe}.
Then the non-linear first order Lie virtual power principle is introduced in~\S~\ref{secpipon}.
It produces models for non linear first order fluids (possibly turbulence), see \S~\ref{exafs1nl},
non-linear first order elasticity, see \S~\ref{secnle1},
and Maxwell type visco-elastic model, see \S~\ref{secvis}.
Then the second order Lie virtual power principle is introduced in \S~\ref{secl2},
to get enriched models for non linear fluids, non linear elasticity,
as well as second order models with some elasticity and some viscosity.
And a quite long appendix is provided since, to the knowledge of the author,
%and as Marsden and Hughes write,
there is no obvious (short) reference in the existing literature concerning covariant objectivity
and Lie derivative for classical mechanics.

\def\point{{p}}
\def\pointt{{p_t}}
\def\pointtau{{p_\tau}}
\def\Point{{P}}
\def\calR{{\cal R}}
\def\vA{{\vec A}}

\section{Notation}
\label{secnot}

We use $\ds\eqdef$ to mean ``defined by''.
Time and space are decoupled (classical mechanic).
The geometric affine space is~$\RRn $, $ n = 1$, $2$ or~$ 3 $.
The associated vector space is also written $\RRn$ (context removes ambiguities),
and is equipped with its usual topology.
And $\RRns=\calL(\RRn;\RR)$ is its dual space (the space of linear forms on~$\RRn$).
A~basis $(\ve_i)_{i=1,...,n}$ is simply denoted~$(\ve_i)$.
An observer defines a referential $\calR = (O, (\ve_i))$ (an origin and a basis,
an origin of time and a timescale being implicit).
Let $ \tu < \td $.
The observer locates the particles of an object $\Obj$
with the mapping, called the motion of~$\Obj$ in~$\calR$,
\begin{equation}
\label{eqtPhi}
\tPhi : 
\left\{\eqalign{
[\tu,\td] \times \Obj & \rar \RRn \cr
(t,\Pobj) & \rar \pointt = \tPhi(t,\Pobj) = O+ \sumin x^i \ve_i,
}\right.
\end{equation}
where $\pointt$ is to position of~$\Pobj$ at~$t$ in~$\calR$.
For $t$ fixed, let $\tPhi_t(\Pobj) := \tPhi (t,\Pobj)$.
The configuration of~$\Obj$ at~$t$ is $\Omegat \eqdef \tPhi_t (\Obj) $.
And let $\bigC := \bigcup_{t\in[\tu,\td]}(\{t\}\times\Omega_t)$, a subset in the standard Newtonian spacetime.

\comment{
An Eulerian tensorial mapping is a mapping
$
\widetilde\eul : 
(t,\pointt) \in \bigC \rar \widetilde\eul(t,\pointt) = ((t,\pointt),\eul(t,\pointt)) \in \bigC\times \calL^r_s(\RRnt)
$
where $\calL^r_s(\RRn) = \calL((\RRnt^*)^r,(\RRnt)^s ; \RR)$ is the set of multilinear forms
acting on $(\RRnt^*)^r\times(\RRnt)^s$
(the mathematical definition of a tensor is recalled in~\S~\ref{secAtco}).
If no confusion arises then $ \widetilde \eul (t, \pointt)$ is abbreviated as $\eul(t,\pointt)$.
For $t$ fixed, let $\eul_t (\pointt) := \eul (t, \pointt)$.
}

%The spaces $ \RR $, $\RRn$ and $\RR\times \RRn$ are provided with the usual topology.
At~$t$, the geometric space $ \RRn $ is written~$\RRnt $
(on a differential manifold the tangent bundle $T\Omegat$ is considered).
Assuming $\tPhi$ is $C^1$ in time,
the Eulerian velocity field $\widetilde\vv : (t,\pointt)\in\bigC \rar ((t,\pointt),\vv(t,\pointt)) \in \bigC \times \RRn$
is defined by
%$\vv$ is the Eulerian vector field defined with $\pointt = \tPhi(t,\Pobj) \in \Omegat$ by
\begin{equation}
\label{eqdefvv}
\vv (t, \pointt) \eqdef \lim_ {h \rar 0} {\tPhi (t {+} h,\Pobj) - \tPhi(t,\Pobj) \over h}
= {\pa \tPhi \over \pa t} (t, \Pobj) \in \RRnt.
\end{equation}
If no confusion arises then $ \widetilde \vv (t, \pointt)$ is abbreviated as $\vv(t,\pointt)$
(and $\vv(t,\pointt)$ is drawn at~$(t,\pointt)$).
Assuming $\tPhi$ is $C^2$ in time,
the Eulerian acceleration field %$\widetilde \vgamma$
is defined with $\pointt = \tPhi (t,\Pobj)$ by
$\vgamma(t,\pointt) = {\pa^2 \tPhi \over \pa t^2}(t,\Pobj)$.

For an $C^1$ Eulerian function $\eul$ (defined on~$\bigC$),
the space differential $d\eul(t,\pointt)$ of~$\eul$ at $(t,\pointt) \in \{t\}\times\Omega_t$
is the differential $d\eul_t(\pointt)$ of~$\eul_t$ at~$\pointt$.
(In differential geometry the tangent map $T\eul_t$ of $\eul_t$ is used,
where $T\eul_t(\pointt) := (\pointt,d\eul_t(\pointt))$,
that is $T\eul_t$ is the ``full notation'' of~$d\eul_t$.)
And its material derivative ${D \eul \over Dt}$ is its derivative along a trajectory,
that is, %at $t$ and $\pointt = \tPhi(t,\Pobj)$, %${D \eul \over Dt}(t,\pointt)$
is the time derivative of the function $t \rar \eul(t,\tPhi(t,\Pobj))$ at $\pointt = \tPhi(t,\Pobj)$;
So, ${D \eul \over Dt}(t,\pointt)
= {\pa \eul \over \pa t}(t,\pointt) + d \eul (t, \pointt). \vv (t, \pointt)$.
\Eg, $\vgamma = {D\vv \over Dt}$.

Let $\tz \in ]\tu,\td[$ be an initial time of computation as set by an observer.
The notations of Marsden and Hughes~\cite{MarsdenHughesMFE} are used:
Capital letters in~$ \Omegatz $ and tiny letters in~$\Omegat$.
And with $t\in [\tu,\td]$, let $\Phitzt : \Omegatz \rar \Omegat$ be defined by
$\Phitzt := \tPhi_t \circ (\tPhi_\tz)^{-1}$, that is
\begin{equation}
\label{eqPhitzt}
\hbox{if}\quad  p_\tz=\Point = \tPhi (\tz, \Pobj) \qthen
\pointt = \Phitzt(\Point) \eqdef  \tPhi (t, \Pobj).
\end{equation}
And the motion relative to the configuration~$\Omegatz$ is
$\Phitz:[\tu, \td] \times \Omegatz \rar \RRn$ defined by
\begin{equation}
\label{eqPhitzt0}
(p_t=)\quad \Phitz(t, \Point) \eqdef \Phitzt(\Point).
\end{equation}
So, $\Phitz(t, \Point) = \tPhi (t, \Pobj) = \pointt$ when $\Point = \tPhi (\tz, \Pobj)$.

The Lagrangian velocity $\vVtz : [\tu,\td]\times \Omegatz \rar \RRnt$ relative to~$\tz$ is defined by
$\vVtz(t,\Point) := {\pa\Phitz \over \pa t}(t,\Point)$.
Let $\vVtzt(\Point):=\vVtz(t,\Point)$.
(The mapping $\vVtzt : \Omegatz \rar \RRnt$ does not define a ``vector field''
but a two points ($\Point$~and~$\pointt$) mapping, see Marsden and Hughes~\cite{MarsdenHughesMFE}
(the vector $\vVtz(t,\Point)$ is drawn at $(t,\pointt)$
since $\vVtz(t,\Point) = \vv(t,\pointt)$ is the tangent vector at~$\pointt$ along
the trajectory $t\rar \tPhi(t,\Pobj)=\Phitz(t, \Point)$).

$\Phitzt$ will be assumed to be a~$ C^2 $ diffeomorphism,
and its differential $\Ftzt := d\Phitzt : \Omegatz \rar \calL(\RRntz;\RRnt)$
is named the (covariant) deformation gradient between~$\tz$ and~$t$.
%As Marsden and Hughes~\cite{MarsdenHughesMFE} point out,
($\Ftzt$ does not define a tensor
but a two points ($\Point$ and~$\pointt$) mapping, see~\cite{MarsdenHughesMFE}.)
Let $\Ftz$ be defined by $ \Ftz(t,\Point) \eqdef \Ftzt(\Point)$.

Let $L(E;F)$ be the set of continuous linear mappings between two vector spaces $E$ and~$F$,
and let $ \dd_G $ (resp. $\dd_g$) be an inner product in~$ \RRntz $ (resp. in~$\RRnt$).
The transpose $ (\Ftzt (\Point))^T \in \calL (\RRnt; \RRntz) $,
relative to $\dd_G$ and~$\dd_g$ is characterized by
$((\Ftzt (\Point))^T. \vw_\pointt, \vW_\Point)_G = (\vw_\pointt, \Ftzt (\Point).\vW_\Point)_g$
for all $\vW_\Point \in \RRntz$ and $\vw_\pointt \in \RRnt$, see~\cite{MarsdenHughesMFE}.
%(If  $\dd_G$ and $\dd_g$ are Euclidean inner products then the transpose $F^T$ does not depend on the choice of $\dd_G$ and~$\dd_g$.)
And the deformation tensor relative to $\tz$, $t$, $\dd_G$ and $\dd_g$ is
$ \Ctzt := (\Ftzt)^T \circ \Ftzt : \Omegatz \rar \calL (\RRntz; \RRntz) $,
written $ C = F^T.F $.

\begin{remark}
\label{remshifter0}
To be able to impose ``the same Euclidean basis at two distinct times $\tz$ and~$t$''
(and on two different tangent spaces)
Marsden and Hughes~\cite{MarsdenHughesMFE} p.~57 use the shifter:
In~$\RRn$ it is the mapping
$ \Stzt: (\Point, \vW_\Point) \in \Omegatz \times \RRntz \rar
\Stzt (\Point, \vW_\Point) = (\pointt, \vw_\pointt) \in \Omegat \times \RRnt $
where $\pointt = \Phitzt(\Point)$ and $ \vw_\pointt \eqdef \vW_\Point$
(parallel displacement in~$\RRn$).
In particular, using ``the same Euclidean basis at~$ \tz $ and~$ t $'' means
using a Euclidean basis~$(\vE_i)$ in~$\RRntz$ and the basis $ (\ve_i) $ in~$ \RRnt $ given by
$ (\pointt, \ve_i) = \Stzt (\Point, \vE_i) $, so $\ve_i=\vE_i$.
\end{remark}

Representation.
Let $\RRnt^* = \calL(\RRnt;\RR)$ be the space of linear functions on~$\RRnt$.
If $(\ve_i)$ is a basis in $\RRnt$,
its dual basis $(e^i)$ is the basis of $\RRnt^*$
made of the linear functions $e^i$ characterized by
$e^i(\ve_j)=\delta^i_j$ (Kronecker symbol).
The $e^i$ being linear, $e^i(\vv)$ is written $e^i.\vv$ for any $\vv\in\RRnt$.
A basis in~$\RRnt$ will be supposed Cartesian to simplify.
At~$\tz$ a basis in~$\RRntz$ will be denoted $(\vE_I)$ and its dual basis~$(E^I)$.
At~$t$ in a referential $\calR=(o_t,(\ve_i))$,
a point $\pointt=\Phitzt(P)$ is located  as the bipoint vector
$\ora{o_t \pointt} = \ora{o_t\Phitzt(\Point)} = \sumin \Phi^i_t (\Point) \ve_i$ in~$\RRnt$,
and the deformation gradient is denoted
\begin{equation}
\label{eqFtzb}
d\Phitzt(\Point) = \Ftzt(\Point)
=  \sumiJn F^i_J(t,\Point)\, \ve_i \otimes E^J ,\quad [F]_{| \vE, \ve} = [F^i_J].
\end{equation}
So, $F^i_J(t,\Point) = d\Phi^i(t,\Point).\vE_J$,
and with Cartesian bases $F^i_J(t,\Point)={\pa\Phi^i \over \pa X^J}(t,\Point)$.
And $[F^i_J (t, \Point)] = [\Ftzt (\Point)]_{| \vE, \ve} = [d\Phitzt(\Point)]_{| \vE, \ve}$
is the Jacobian matrix of~$\Phitzt$ at~$\Point$ relatively to the bases~$(\vE_I)$ and~$(\ve_i)$.
Since $\Phitzt$ is supposed to be a diffeomorphism
we have $d\Phitzt(P)^{-1} = (d\Phitzt)^{-1}(\pointt)$ when $\pointt = \Phitzt(P)$,
and we will denote $\Htzt(\pointt):=(d\Phitzt)^{-1}(\pointt)$. And with the above bases:
\begin{equation}
\label{eqFtzbh}
\Htzt(\pointt) = \Htz(t,\pointt)
= \sum_{I,j=1}^n H^I_j(t,\pointt) \vE_I \otimes e^j , \quad [H]_{|\ve,\vE} = [H^I_j].
\end{equation}

\section{Classical virtual power principle}
\label{seccvpp}

\def\paOmegat{{\pa\Omegat}}

We use Germain's setting, see~\cite{GermainAPPV1973},
with no volumic double forces to simplify.
Let $\dd_g$ be a Euclidean inner product in~$\RRnt$.
Let $\calV $ be a vector space of ``admissible velocity fields'' (sufficiently regular vector fields
for the mathematical expressions to be meaningful).
The virtual power principle connects, at all times~$t$,
three linear functionals $\calV \rar \RR$:
The virtual power of external forces
\begin{equation}
\label{eqcalPe}
\calPe(\vv)
= \int_{\Omegat} (\vf,\vv)_g \,d\Omega
+ \int_{\paOmegat} (\vg,\vv)_g \,d\sigma
\end{equation}
where $\vf $ and $\vg $ are given vector fields,
the virtual power of mass-acceleration
\begin{equation}
\label{eqcalPa}
\calPa(\vv) 
\eqdef \int_{\Omegat} \rho (\vgamma,\vv)_g \,d\Omega
\end{equation}
where $\rho$ is the mass density and $\vgamma = {D\vw\over Dt}$
is the eulerian acceleration undergone by a material with velocity~$\vw$,
and the virtual power of internal forces
\begin{equation}
\label{eqcalPG}
\calPi(\vv) = \int_{\Omegat} \pint(\vv)\,d\Omega
\end{equation}
where $\pint $ is a linear form of~$\vv $ and its differentials.
And the principle is: For any $\vv\in \calV$,
\begin{equation}
\label{eqppv}
\calPi(\vv) + \calPe(\vv) = \calPa(\vv). 
\end{equation}

\comment{
\begin{example}
Incompressible Stokes fluid animated with a velocity~$\vw$: In a Galilean Euclidean setting,
$\pint(\vv) = \uusigma : d\vv$ with
%\begin{equation}
%\label{eqsm10}
$\uusigma = 2\mu\,{d\vw + d\vw^T \over 2} - pI$,
%\end{equation}
where $p$ is the pressure, $\mu$ the viscosity coefficient,
$\calPi(\vv) = \int_\Omegat \uusigma : d\vv \, d\Omega$ for all $\vv\in\calV $,
and where $\uusigma : d\vv := [\uusigma]_{|\ve}:[d\vv]_{|\ve}$ is the double matrix contraction
of $[\uusigma]_{|\ve}$ with $[d\vv]_{|\ve}$, see~\eref{eqttg}.
\end{example}
}

\def\vazzj{{\va_{00j}}}
\def\vazui{{\va_{01i}}}
\def\vazuj{{\va_{01j}}}
\def\vauj{{\va_{1j}}}
\def\vazuk{{\va_{01k}}}
\def\vauzi{{\va_{10i}}}
\def\vauzj{{\va_{10j}}}
\def\vauzk{{\va_{10k}}}
\def\vauuj{{\va_{11j}}}
\def\alphazzi{{\alpha_{00}^i}}
\def\alphazzj{{\alpha_{00}^j}}
\def\alphazui{{\alpha_{01}^i}}
\def\alphazuj{{\alpha_{01}^j}}
\def\alphauj{{\alpha_{1}^j}}
\def\alphaui{{\alpha_{1}^i}}
\def\alphazuk{{\alpha_{01}^k}}
\def\alphauzj{{\alpha_{10}^j}}
\def\alphauzi{{\alpha_{10}^i}}
\def\alphauuj{{\alpha_{11}^j}}
\def\alphauui{{\alpha_{11}^i}}
\def\alphadzj{{\alpha_{20}^j}}
\def\alphadzi{{\alpha_{20}^i}}

\section{Virtual power principle with Lie derivatives}
\label{secppvl}

\def\pres{{pres}}

The virtual power of external forces \eref{eqcalPe} and of mass-acceleration \eref{eqcalPa}
are used; And the ``virtual power of pressure''
\begin{equation}
\label{eqcalPpres}
\calPpres(\vv)
= \int_\Omegat p\;\dvg\vv\;d\Omegat
\end{equation}
is introduced, where $ p $ is a differentiable function,
and $\dvg\vv$ is the divergence of the velocity field~$\vv$. % (the trace of~$d\vv$).
(The virtual power of pressure will enable a simple formulation for the virtual power of internal forces
as is classically done for Maxwell visco-elastic material, see~\S~\ref{secvis}.
Also see~\S~\ref{secsel}.)

And the virtual power of internal forces \eref{eqcalPG} will be used
with $\pint$ (non linear in general) given with Lie derivatives as described in~\S~\ref{secpfei1} and following:
This is the main purpose of this manuscript.

Then the principle of virtual power reads:
for any admissible vector field~$\vv $,
\begin{equation}
\label{eqppvlie}
\calPi(\vv) + \calP_\pres(\vv) + \calPe(\vv) = \calPa(\vv) .
\end{equation}

\section {Lie virtual power of internal forces: Linear first order}
\label{secpfei1}

(The Lie derivatives in classical mechanics are described in \S~\ref{secdlfei}.)

\subsection {First Order Formulation}

First-order conjecture:
1-~A material, occupying at~$t$ a domain $\Omegat$ in~$\RRn$, can be characterized
by $3n$ vector fields $\vazzj, \vazuj, \vauzj$
and/or $ 3n $ one-forms $ \alphazzj, \alphazuj, \alphauzj$.
2-~At~$t$, in a referential $\calR=(O,(\ve_i))$ and with $(e^i)$ the dual basis of~$(\ve_i)$,
the measured density of power of internal forces depends on an admissible virtual velocity field~$\vv$ acting on the material,
and reads
\begin{equation}
\label{eq10}
\pint_1(\vv)
=  \sumjn e^j.\vazzj  + \calL_\vv e^j.\vazuj + e^j.\calL_\vv \vauzj
+  \alphazzj.\ve_j + \alphazuj.\calL_\vv \ve_j + \calL_\vv \alphauzj.\ve_j,
\end{equation}
where $\calL_\vv \va = {\pa \va\over \pa t} + d\va.\vv - d\vv.\va$
is the Lie derivative of a vector field~$\va$, see~\eref{eqdefLiev3},
and $\calL_\vv \alpha = {\pa \alpha \over \pa t} + d \alpha. \vv + \alpha.d \vv$
is the Lie derivative of a one-form~$\alpha$, see~\eref{eqpfalpha}.
(For special materials the number of vector fields and/or one-forms can be chosen to be greater than $n$ the dimension of~$\RRn$.)

\subsection {Galilean setting}
\label{seccg}

\def\uutauu{{\uutau_1}}

In a Galilean referential and with a Cartesian basis $(\ve_i)$,
we have $\calL_\vv \ve_j = -d\vv.\ve_j$ and $\calL_\vv e^j = e^j.d\vv$,
and \eref{eq10} reduces to
$
\pint_1(\vv)
= \sumjn e^j.\vazzj + e^j.d\vv.\vazuj + e^j.({\pa \vauzj\over \pa t} + d\vauzj.\vv - d\vv.\vauzj)
+ \alphazzj.\ve_j - \alphazuj.d\vv.\ve_j
+ ({\pa \alphazuj \over \pa t} + d \alphazuj. \vv + \alphazuj.d \vv).\ve_j
$.
And the virtual power of the internal forces is assumed to vanish whenever $d\vv =0$
(Galilean referential). Thus we are left with
$\pint_1(\vv)
= \sumjn e^j.d\vv.\vazuj - e^j.d\vv.\vauzj - \alphazuj.d\vv.\ve_j + \alphazuj.d \vv.\ve_j$,
that is,
\begin{equation}
\label{eqh1}
\pint_1(\vv)= -\uutauu \odd d\vv,
\quad
-\uutauu = \sumjn \vauj\otimes e^j + \ve_j\otimes \alphauj ,
\quad
\left\{\eqalign{
& \vauj = \vazuj-\vauzj, \cr
& \alphauj = \alphazuj-\alphazuj,
}\right.
\end{equation}
where $\odd$ is the double objective contraction
between the ${1\choose1}$ tensors $\uutau_1$ and~$d\vv$, see~\eref{eqToddS2}. %-\eref{eqToddS2c}.

\subsection {Galilean Euclidean setting}
\label{seccge}

In the previous Galilean setting, %at~$t$
the basis $(\ve_i)$ is chosen to be Euclidean.
Let $\dd_g$ be the associated inner product.
Then the transposed $d\vv^T$ of~$d\vv$ is defined (by $(d\vv^T.\vu,\vw)_g = (d\vv.\vw,\vu)_g$
for any $\vu,\vw\in\RRnt$). %, that is, $\dd_g = \sumin e^i\otimes e^i$.
The internal power being independent of a rigid motion,
i.e independent of~$ {d \vv - d \vv^T \over 2} $, \eref{eqh1} gives
\begin{equation}
\label{eqh18}
\pint_1(\vv) =  -\uutauu \odd {d\vv + d\vv^T \over 2}
 =  -\uusigma \odd {d\vv + d\vv^T \over 2} 
 =  -\uusigma \odd d\vv,
\quad \uusigma = {\uutauu +\uutauu^T \over 2}, 
\end{equation}
which is a classical expression. %, except for the pressure term taken into account with~\eref{eqcalPpres}.

\section {Stokes Fluid}
\label{exafs}

The ``pressure forces'' being taken into account in the virtual power of pressure, see~\eref{eqcalPpres},
we look at the virtual internal power of ``viscous forces''.
Consider a fluid animated with a movement $\tPhi$, see~\eref{eqtPhi},
let $p(\tau) = \tPhi(\tau,\Pobj)$,
and let $\vw(\tau,p(\tau)) = {\pa\tPhi \over \pa \tau}(\tau,\Pobj)$ be its Eulerian velocity, see~\eref{eqdefvv}.
Then, with $t\in]t_1,t_2[$ and $\tau$ in the vicinity of~$t$, consider the associated movement $\Phi^t_\tau$, see~\eref{eqPhitzt},
and its deformation gradient $\Fttau = d\Phi^t_\tau$ between $t$ and~$\tau$.
Since
${\pa\Phi^t \over \pa \tau}(\tau,\pointt) = \vw(t,\Phi^t(\tau,\pointt))$
we have ${\pa F^t \over \pa \tau}(\tau,\pointt) = d\vw(t,p(\tau)).F^t(\tau,\pointt)$.

%\noindent
{\bf Eventuality 1.} Vector fields $\vauj$ characterize the fluid,
and the virtual power of internal forces results from their transport by the flow.
Then %, $\vv$ being a virtual admissible vector field,
\eref{eqh1} gives (Galilean setting)
\begin{equation}
\label{eqfev2b}
\pint_1(\vv) = \sumjn e^j. \calL_ \vv \vauj = -\uutauu \odd d\vv,
\quad
\uutauu = -\sumjn \vauj \otimes e^j .
\end{equation}
Then consider that a Newtonian fluid (which is isotropic) is characterized
by the $n$ vector fields given by
$\vauj(\tau,\pointtau) := -2\mu\,{\pa F^t \over \pa \tau}(\tau,\pointt).\ve_j$ when $\tau=t$ (no memory),
that is,
\begin{equation}
\label{eqfev2c}
\vauj(t,\pointt) = -2\mu\,d\vw(t,\pointt).\ve_j.
\end{equation}
(The matrix   $ [\uutauu]_{|\ve}  = 2\mu\, [d\vw]_{|\ve}$ is made of the columns
$ -[\vauj] _ {|\ve} = 2 \mu \, [{\pa \vw \over \pa x^j}]_{|\ve} $.)
Thus, in a Galilean Euclidean setting, 
we get
$[\uusigma]_{|\ve} %= {\uutauu + \uutauu^T \over 2}
= 2\mu {[d\vw]_{|\ve} + [d\vw]_{|\ve}^T \over 2}$, see~\eref{eqh18}, which is the classical model.

\Eg\ if $\vauj=\vauzj$, that is, if $\pint_1(\vv) = \sumjn e^j.\calL_\vv \vauzj$, see~\eref{eq10} and~\eref{eqh1},
then the power is measured by the observer with the projections~$e^j$.
(And if $\vauj=\vazuj$, that is if $\pint_1(\vv) = \sumjn \vazuj.\calL_\vv e^j$,
then the power is measured with the~$e^j$ immersed in the flow: will be used with the non linear approach.)

%\noindent
{\bf Eventuality 2.}
One-forms characterize the fluid,
and their transport by a flow~$ \vv $ give the internal power.
In particular in a Galilean Euclidean setting with a Euclidean basis $ (\ve_i) $ in~$ \RRnt $,
\eref{eqh1} gives
$\pint_1(\vv) =-\uutauu \odd d\vv$ where $\uutauu = -\sumin \ve_i \otimes \alphaui$.
Here $ \alphaui = -2 \mu \, e^j.d\vw = -2 \mu \, dw^i $, and
$[\uutauu]_{|\ve}$ is the matrix made of the lines $-[\alphauzi]_{|\ve} = 2 \mu [dw^i]_{|\ve}$.

\section{Linear elasticity}
\label{secsel0}

\def\alphauitz{{\alpha^{i}_{1\tz}}}
\def\alphajtz{{\alpha^{j}_\tz}}
\def\vajtz{{\va_{j\tz}}}

%The ``pressure forces'' being taken into account in the virtual power of pressure, see~\eref{eqcalPpres}, we look at the virtual internal power without the ``trace part'' of the stress tensor, see~\eref{equs1p} or~\eref{equs1p2}, and \eref{equs2} or~\eref{equs2pb}.
Consider an elastic solid animated with a movement $\tPhi$, see~\eref{eqtPhi},
let $\tz\in]t_1,t_2[$, let $\Phitz$ be the associated movement,
see~\eref{eqPhitzt0} and~\eref{eqPhitzt},
and let $\Ftzt = d\Phitzt$ be the covariant deformation gradient between $\tz$ and~$t$.

{\bf Eventuality 1.}
One-forms characterize the elastic material (Germain's point of view see~\cite{GermainAPPV1973a}),
and their transport by the flow gives the internal power.
Then~\eref{eqh1} gives (Galilean setting)
\begin{equation}
\label{eqevs1b}
\pint_1(\vv)
= \sumin \calL_ \vv \alphaui. \ve_i
= -\uutauu \odd d\vv, \quad \uutauu = -\sumin \ve_i\otimes \alphaui .
\end{equation}
Then consider that an isotropic elastic solid is characterized
by $n$ one-forms $\alphaui$ on~$\Omegat$ that are push-forwards of one-forms $\alphauitz$ on~$\Omegatz$,
see~\eref{eqpbalpha}, that is,
with $\pointt = \Phitzt(\Point)$ and $\Htzt(\pointt) = (\Ftzt (\Point))^{- 1}$, % = \Htz(t,\pointt)$,
\begin{equation}
\label{eqevs11}
\alphaui(t,\pointt) = \alphauitz(\Point).\Htzt(\pointt).
\end{equation}
Then $\uutauu(t,\pointt) = -\sumin (\ve_i \otimes \alphauitz(\Point)).\Htz(t,\pointt)$;
And with $(\vE_i)$ a Cartesian basis in~$\RRntz$,
choose $\alphauitz=E^i$ (isotropic material), so that, with~\eref{eqFtzbh},
$\uutauu(t,\pointt)
= -\sumin (\ve_i \otimes E^i).\Htz(t,\pointt)
= - \sum_{i,k,j=1}^n H^k_j(t,\pointt) (\ve_i \otimes E^i).(\vE_k \otimes e^j)
= - \sum_{i,j=1}^n H^i_j(t,\pointt) \ve_i \otimes e^j
$, that is,
\begin{equation}
\label{eqevs1c}
\uutauu(t,\pointt)
%= -\sumin (\ve_i \otimes E^i).\Htz(t,\pointt)
= -\sum_{i,j=1}^n H^i_j(t,\pointt) \ve_i \otimes e^j,\qand
[\uutauu]_{|\ve} = -2\tmu [\Htz]_{|\ve,\vE}.
\end{equation}
So, in a Galilean Euclidean setting, with $(\vE_i)$ a Euclidean basis, we get
$[\uusigma]_{|\ve} = -2\tmu {[\Htz]_{|\ve,\vE}+[\Htz]_{|\ve,\vE}^T \over 2}$, see~\eref{eqh18},
classical expression for small displacements, except for the trace part, see \S~\ref{secsel}.

\Eg\ if $\alphauj=\alphauzj$, that is if $\pint_1(\vv) = \sumjn \calL_\vv \alphauzj.\ve_j$,
then the power is measured by the observer with the Euclidean projections along the~$\ve_j$.
(And if $\alphauj=\alphazuj$, that is if $\pint_1(\vv) = \sumjn \alphazuj.\calL_\vv \ve_j$,
then the power is measured with the $\ve_j$ immersed in the flow; It will be used for visco-elasticity,
see~\eref{eqTM0101m}.)

{\bf Eventuality 2.}
Vector fields characterize the solid,
and their transports give the internal power.
In particular in a Galilean Euclidean setting with a Euclidean basis $ (\ve_i) $ in~$ \RRnt $,
\eref{eqh1} gives
\begin{equation}
\pint_1(\vv) = -\uutauu \odd d\vv, \quad \uutauu = -\sumjn \vauj \otimes e^j.
\end{equation}
Then suppose that the $ \vauj $ are the push-forwards of vector fields $\vajtz (\Point)$ see~\eref{eqvwvWb}, that is, with $\pointt = \Phitzt(\Point)$,
$\vauj(t,\pointt) = \Ftzt(\Point).\vajtz(\Point)$.
Then $\uutauu(t,\pointt) = -\Ftz(t,\Point).\sumjn \vajtz(\Point)\otimes e^j$.
With a Euclidean base $ (\vE_i) $ in~$ \RRntz $, then for isotropic homogeneous elasticity
we may choose  $ \vajtz = -2 \tmu \vE_j $, then, see~\eref{eqFtzb}:
\begin{equation}
\label{eqev2}
\uutauu(t,\pointt) = \Ftz(t,\Point).\sumjn \vE_j\otimes e^j,\qand
[\uutauu]_{|\ve} = 2\tmu [\Ftz]_{|\vE,\ve}. %\qand [\uusigma] = 2\tmu {[\Ftz]+[\Ftz]^T \over 2},
\end{equation}
Thus, in a Galilean Euclidean setting, with $(\vE_i)$ a Euclidean basis, we get
$[\uusigma] = 2\tmu {[\Ftz]+[\Ftz]^T \over 2}$,
classical expression for small displacements, except for the trace part, see \S~\ref{secsel}.

\section{Fluids vs solids, and hysteresis}
\label{secfvs}

For a Stokes fluid, a modelization with vector fields can be considered, 
see~\eref{eqfev2b}.
And for an elastic solid, a modelization with one-forms can be considered,
see~\eref{eqevs1b}.
Thus fluids would be differentiated from solids in a non-algebraic way.
And ``mixed linear materials'' could be considered with
\begin{equation}
\label{eqh}
\pint_1(\vv)
= c_1 \sumjn e^j.\calL_\vv \vauzj + c_2\sumin \calL_\vv \alphauzi.\ve_i
= - c_1\, \uutauu \odd d\vv - c_2\, \uukappa_1 \odd d\vv, 
\end{equation}
the last equality in a Cartesian setting
where 
$\uutauu = - \sumjn \vauzj \otimes e^j$
characterizes the fluid part,
$\uukappa_1 = - \sumin \ve_i \otimes \alphauzi$
characterizes the solid part,
and $ c_1, c_2 \in \RR $.
This could model simple visco-elasticity or hysteresis.

\section{About hyper-elasticity}
\label{secahe}

\def\calW{{\cal W}}
\def\calWtz{{\calW^\tz}}
\def\calWtzt{{\calW^\tz_t}}
\def\PK{{P\!\! K}}
\def\hPK{{\widehat\PK}}
\def\hPKtz{{\widehat\PK_\tz}}

Hyper-elasticity aims to find a ``stored energy function'' see Marsden and Hughes~\cite{MarsdenHughesMFE}
(``\'energie volumique des d\'eformations \'elastiques'' see Germain~\cite{GermainX1984}).
The classical starting point is a Euclidean setting and the Piola--Kirchhoff tensor
$\PK = J \uusigma.F^{-T} $, simplified notation of
$\PK^\tz_t(\Point) = J^\tz_t(\Point) \uusigma (t, \pointt) .(\Ftzt (\Point))^{-T} $,
where $\pointt = \Phitzt (\Point)$,
$ J = \det (F) $,
$ F^{-T} $ is the inverse of the transpose,
$ \uusigma $ is the Cauchy stress tensor at $(t,\pointt)$
built from the Cauchy stress vector %$\vec T$ 
at $(t,\pointt)$ and
 the unit orthonormal vector $\vn(t,\pointt)$ to~$\paOmegat$ (depends on the metric).
Thus $ \PK $ measures the force ``per unit of undistorted area'' (by change of variables
in the integrals).
And the material is said to be hyper-elastic if $ \PK^\tz(t, \Point) = \hPK^\tz (t, \Ftz (t, \Point)) $
and if there exists a function~$\calW^\tz$ s.t.,
with~\eref{eqFtzb}, $[\hPK^\tz]  =  [{\pa \calW^\tz \over \pa F^i_J}]$.
This derivation in terms of the components of the two-point tensor~$F$
is quite intriguing (see~\cite{MarsdenHughesMFE}).
%moreover when it is noticed that $\hPKtz$ is not symmetric (cannot be since $\hPK^\tz_t : \RRntz \rar \RRnt$ is not an endomorphism, see~\cite{MarsdenHughesMFE}).

With the Lie derivative approach,
and the introduction of the virtual power of pressure, see~\eref{eqcalPpres},
the "hyper-elastic potential"  for isotropic homogeneous elasticity can be considered to be simply
$\calWtzt := 2\tmu\Phitzt$:
Then $d\calWtzt = 2 \tmu \Ftzt$, % : \Omegatz \rar\calL(\RRntz;\RRnt)$,
and then $[\uutauu(t,\pointt)]_{|\ve}
= [d \calWtz(t,\Point)]_{|\vE,\ve}
= 2 \tmu [\Ftz(t,\Point)]_{|\vE,\ve}$, see~\eref{eqev2}.
But the preference may go to
\begin{equation}
\label{eqWhe}
\calWtzt := -2\tmu (\Phitzt)^{-1} : \Omegat \rar \Omegatz
\end{equation}
%(the potential at $(t,\pointt)$),
so that $d\calWtzt(\pointt) = -2\tmu\Htzt(\pointt)$, % : \Omegat \rar\calL(\RRnt;\RRntz)$ 
and 
$[\uutauu(t,.)]_{|\ve} = [d \calWtzt]_{|\ve,\vE} = -2 \tmu [\Htzt]_{|\ve,\vE}$, see~\eref{eqevs1c}. %that is $\uutau_1(t,\pointt) = d\calWtz(t,\pointt)$.
Here $\uutauu(t,.)$ and~$d\calWtzt$, in~\eref{eqWhe}, are defined at~$\pointt$,
and $\calWtzt$ refers to the past (values at~$\tz$ from~$\Omegat$) which is the usual approach of Galilean or general relativity.

\section{Lie virtual power of internal forces: Non linear first order}
\label{secpipon}

\def\uutauuu{{\uutau_{11}}}

\eref{eq10} is enriched:
1'-~The material can be characterized by the preceding fields and by
$n$ vector fields $\vauuj$ on~$\Omegat$ and/or $n$ one-forms $\alphauuj$ on~$\Omegat$.
2-~For a velocity field~$\vv$, at~$t$, we get the density of the power of internal forces (first order nonlinear)
\begin{equation}
\label{eq1n}
\pint_{1n}(\vv)
=  \pint_1(\vv) 
+ \sumjn \calL_\vv \alphauuj.\calL_\vv \ve_j
+ \calL_\vv e^j.\calL_\vv \vauuj.
\end{equation}

In a Galilean setting,
the assumption of zero internal power if $ d \vv = 0 $ gives
with~\eref{eqh1}:
$$
\pint_{1n}(\vv)
=  -\uutauu \odd d\vv
+ ({\pa\alphauuj \over \pa t} + d\alphauuj.\vv + \alphauuj.d\vv).(-d\vv.\ve_j)
+ e^j.d\vv.({\pa\vauuj \over \pa t} + d\vauuj.\vv - d\vv.\vauuj)
.
$$
In~\eref{eq1n}, in addition to $\pint_1(\vv) = -\uutauu \odd d \vv $, it appears:

the unsteady linear term
$-{\pa \alphauuj \over \pa t}.d\vv.\ve_j + e^j.d\vv.{\pa\vauuj \over \pa t}$,

the non-linear term
$-(d\alphauuj.\vv).d\vv.\ve_j + e^j.d\vv.(d\vauuj.\vv)$, non-linear in $\vv$ and~$d\vv$,

and the non-linear term
$-\alphauuj.d\vv.d\vv.\ve_j - e^j.d\vv.d\vv.\vauuj$, non-linear in~$d\vv.d\vv$.

\section{Non Newtonian fluids 1}
\label{exafs1nl}

$2n$ vector fields $\vauj$ and $\vauuj$ characterize a non linear first order fluid
(non Newtonian, turbulence type model?).
Then, with~\eref{eqfev2b} and $\uutauu = -\sumjn \vauj \otimes e^j$,
\begin{equation}
\label{eqnonlt}
\pint_{1n}(\vv)
= -\uutauu \odd d\vv
+ \sumjn ({\pa\vauuj \over \pa t} \otimes e^j + (d\vauuj.\vv)\otimes e^j) \odd d\vv
- (\vauuj \otimes e^j) \odd (d\vv.d\vv).
\end{equation}
And (for matrix computation) with $\uutauuu = -\sumjn \vauuj \otimes e^j$ we get
\begin{equation}
\label{eqnonlt2}
\eqalign{
\pint_{1n}(\vv)
= & -\uutauu \odd d\vv - ({\pa\uutauuu \over \pa t} + d\uutauuu.\vv - d\vv.\uutauuu) \odd d\vv.
}
\end{equation}
(The expression in parentheses, that is ${\pa\uutauuu \over \pa t} + d\uutauuu.\vv - d\vv.\uutauuu$,
is not a Lie derivative of $\uutauuu$ see~\eref{eqdl}:
Here it is the Lie derivative of the vector fields~$ \vauuj $ that are considered.)

%A simple non linear first order fluid model (non Newtonian, turbulence type model?) is \eg\ given with $ \vauuj = \eta \vauj $, that is, with $ \uutauuu = \eta \uutauu$.

\section{Non linear elasticity 1}
\label{secnle1}

$2n$ one-forms $\alphauj$ and $\alphauuj$ characterize non linear first order elasticity.
Then, with~\eref{eqevs1b} and $\uutauu = -\sumjn \ve_j \otimes \alphauj$, 
\begin{equation}
\label{eqnlela1}
\pint_{1n}(\vv)
= -\uutauu \odd d\vv
- \sumjn (\ve_j\otimes ({\pa \alphauuj \over \pa t} + d\alphauuj.\vv + \alphauuj.d\vv)) \odd d\vv.
%+(\ve_j \otimes \alphauuj) \odd (d\vv.d\vv).
\end{equation}
And (for matrix computation), with $\uutauuu = \sumjn \ve_j \otimes \alphauuj$ we get
\begin{equation}
\label{eqpint1nge2}
\eqalign{
\pint_{1n}(\vv)
= & -\uutauu \odd d\vv - ({\pa\uutauuu \over \pa t} + d\uutauuu.\vv + \uutauuu.d\vv) \odd d\vv.
}
\end{equation}
(The expression in parentheses, that is ${\pa\uutauuu \over \pa t} + d\uutauuu.\vv + \uutauuu.d\vv$,
is not a Lie derivative of $ \uutauuu$ see~\eref{eqdl}:
Here it is the Lie derivative of the one-forms~$ \alphauui $ that are considered.)

\section{Visco-elastic fluids 1}
\label{secvis}

Galilean Euclidean setting. 
The classical model for a visco-elastic fluids of Maxwell reads
\begin{equation}
\label{eqTM010}
\uutau + \lambda \calL_\vv \uutau = \mu {d\vv + d\vv^T \over 2},
\end{equation}
where $\uutau$ is a stress tensor, $\vv$ the velocity field of the material,
$\lambda$ an elasticity constant, $ \mu $ a viscosity constant,
and $\calL_\vv \uutau$ a Lie derivative.
And the pressure~$p$ is classically  introduced with~\eref{eqcalPpres}.
In~\eref{eqTM010}, the derivative of Lie $ \calL_ \vv \uutau $
is considered with the velocity $\vv$ of the material (not a virtual velocity),
and can the Jaumann or the upper-convected or the lower-convected Lie derivative, see~\eref{eqdl};
Or can be a linear combination of such tensors to improve the numerical results
(which otherwise are not convincing),
even if such a linear combination is absurd as far as objectivity is concerned
(addition of vector fields and one-forms).

A possible covariant model could be, with $\vv$ a virtual velocity,
\begin{equation}
\label{eqTM0101m}
\pint_{1n}(\vv)
=  \sumjn e^j.\calL_\vv\vauj + \sumjn \calL_\vv \alphauuj.\calL_\vv \ve_j 
=  -\uutauu \odd d\vv - ({\pa\uutauuu \over \pa t} + d\uutau.\vv + \uutauuu.d\vv) \odd d\vv ,
\end{equation}
obtained by using the Lie derivatives of vector fields for the fluid part and
the Lie derivative of one forms for the elastic part, see~\eref{eqfev2b} and~\eref{eqevs1b},
the last equality being given in a Galilean setting with
$\uutauu = -\sumjn \vauj \otimes e^j$, see~\eref{eqfev2b}, and 
$\uutauuu = \sumjn \ve_j \otimes \alphauui$, see the last term in~\eref{eqpint1nge2}.

\section{Lie virtual power of internal forces: Second order}
\label{secl2}

\def\vazdj{{\va_{02j}}}
\def\vaudj{{\va_{12j}}}
\def\vadzj{{\va_{20j}}}
\def\vaduj{{\va_{21j}}}
\def\vaddj{{\va_{22j}}}
\def\alphazdj{{\alpha_{02}^j}}
\def\alphazdi{{\alpha_{02}^i}}
\def\alphaudj{{\alpha_{12}^j}}
\def\alphadzj{{\alpha_{20}^j}}
\def\alphaduj{{\alpha_{21}^j}}
\def\alphaddj{{\alpha_{22}^j}}
\def\uukappadz{{\uukappa_{20}}}

%\subsection{Formulation}

\eref{eq1n} is enriched with vector fields $\va_{xyj}$ and one-forms $\alpha^j_{xy}$
and the second ordre Lie derivatives

$\alphazdj.\calL_\vv (\calL_\vv \ve_j)$,
$\calL_\vv (\calL_\vv \alphadzj).\ve_j$,
$\calL_\vv \alphaudj.\calL_\vv (\calL_\vv \ve_j)$,
$\calL_\vv (\calL_\vv \alphaduj).\calL_\vv\ve_j$,
$\calL_\vv(\calL_\vv \alphaddj).\calL_\vv (\calL_\vv \ve_j)$,
and

$e^j.\calL_\vv (\calL_\vv \vazdj)$,
$\calL_\vv (\calL_\vv e^j).\vadzj$,
$\calL_\vv e^j.\calL_\vv (\calL_\vv \vaudj)$,
$\calL_\vv (\calL_\vv e^j).\calL_\vv\vaduj$,
$\calL_\vv(\calL_\vv e^j).\calL_\vv (\calL_\vv \vaddj)$.
\\
And, with~\eref{eq1n}, the measured power density of the internal forces reads:
\begin{equation}
\eqalign{
\pint_2(\vv)
= & \pint_{1n}(\vv) + \sum \hbox{the above second order terms}.\cr
}
\end{equation}
The second order Lie derivative of a vector field~$\vw $ is
$\calL_\vv (\calL_\vv \vw)
=  {\pa ^2 \vw \over\pa t^2} + 2 d{\pa \vw \over \pa t}.\vv - 2d\vv.{\pa \vw \over \pa t}
+ d\vw.{\pa \vv \over \pa t} - {\pa d\vv \over \pa t}.\vw
+ (d^2\vw.\vv).\vv + d\vw.d\vv.\vv  - 2 d\vv.d\vw.\vv
- (d^2\vv.\vv).\vw  + d\vv.d\vv.\vw$,
computed with~\eref{eqdefLiev3}.
And the second order Lie derivative of a one-form~$\alpha $ is
$
\calL_\vv (\calL_\vv\alpha)
=  {\pa ^2 \alpha \over \pa t^2} + 2 d{\pa \alpha \over \pa t}.\vv
+ 2 {\pa\alpha \over \pa t}.d\vv + d\alpha.{\pa \vv\over \pa t} + \alpha.{\pa d\vv \over \pa t}
+ d^2\alpha(\vv,\vv) + d\alpha.(d\vv.\vv) + 2 (d\alpha.\vv).d\vv
+ \alpha.(d^2\vv.\vv) + (\alpha. d\vv).d\vv$,
computed with~\eref{eqpbalpha}.

(Generalization: Higher order Lie derivative can also be considered.)

\section{Non Newtonian fluids 2}
\label{sectur2}

\Eg\ \eref{eqnonlt} is enriched to get \eg\ (``one pure second order fluids'')
\begin{equation}
\pint_2(\vv) =  \sumjn e^j.\calL_\vv \vauj
+ \calL_\vv e^j.\calL_\vv \vauuj
+  e^j.\calL_\vv (\calL_\vv \vadzj).
\end{equation}
(And for a ``pure fluid'' any other Lie derivative of vector fields can be added.)

Galilean setting:
The internal power vanishing if~$d\vv=0$ and ${\pa \vv \over \pa t} = 0$,
we get 
\begin{equation}
\eqalign{
\pint_2(\vv)
= & \pint_{1n}(\vv) 
-\sumjn 2e^j.d\vv.{\pa \vadzj \over \pa t}
 + e^j.d\vadzj.{\pa \vv \over \pa t} - e^j.{\pa d\vv \over \pa t}.\vadzj \cr
 & + e^j.d\vadzj.d\vv.\vv  - 2 e^j.d\vv.d\vadzj.\vv
   - e^j.(d^2\vv.\vv).\vadzj  + e^j.d\vv.d\vv.\vadzj, \cr
}
\end{equation}
which is a ``non linear second gradient'' (non Newtonian) expression.
With $\uutau = \sumjn \vauj \otimes e^j$
and $\uukappa_2 = \sumjn \vadzj \otimes e^j$
we obtain (for matrix computation with the generic notation ${DT\over Dt} = {\pa T \over\pa t} + dT.\vv$):
\begin{equation}
\pint_2(\vv)
=  \pint_{1n} (\vv)
 - \Bigl(\uukappa_2 \odd 
({\pa d\vv \over \pa t} + d^2\vv.\vv - d\vv.d\vv)
+ d\uukappa_2.{D \vv \over D t}
 +2{D \uukappa_2 \over D t} \odd d\vv \Bigr).
\end{equation}

\section{Elasticity 2}
\label{secusdasd}

\Eg\ \eref{eqnlela1} is enriched to get \eg\ (``a pure second order elastic material'')
\begin{equation}
\pint_2(\vv)
= \sumin \calL_\vv \alphaui.\ve_i
+ \calL_\vv \alphauuj.\calL_\vv \ve_j
+ \calL_\vv (\calL_\vv \alphadzi).\ve_i.
\end{equation}
(And for a ``pure elastic material'' any other Lie derivative of one-forms can be added.)

Galilean setting:
The internal power vanishing if~$d\vv=0$ and ${\pa \vv \over \pa t} = 0$,
we get 
\begin{equation}
\eqalign{
\pint_2(\vv)
= & \pint_{1n}(\vv) 
+ \sumjn 2 {\pa\alphadzi \over \pa t}.d\vv.\ve_i
+  d\alphadzi.{\pa \vv\over \pa t}.\ve_i
+ \alphadzi.{\pa d\vv \over \pa t}.\ve_i \cr
&+ d\alphadzi.(d\vv.\vv).\ve_i + 2 (d\alphadzi.\vv).d\vv.\ve_i
+ \alphadzi.(d^2\vv.\vv).\ve_i + (\alphadzi. d\vv).d\vv.\ve_i, \cr
}
\end{equation}
which is a ``non linear second gradient'' expression.
With 
$\uukappa_2 = -\sumin \ve_i \otimes \alphadzi$,
we get (for matrix computation):
\begin{equation}
\pint_2(\vv)
=  \pint_{1n}(\vv)
-\Bigl(\uukappa_2 \odd
({\pa d\vv \over \pa t} + d^2\vv.\vv + d\vv.d\vv)
+ d\uukappa_2.{D \vv \over D t}
+ 2{D \uukappa_2 \over D t}\odd d\vv \Bigr).
\end{equation}
With the Eulerian acceleration $ \vgamma = {D\vv \over Dt} = {\pa \vv \over \pa t} + d \vv. \vv $,
that gives $ d \vgamma = {\pa (d \vv) \over \pa t} + d^2 \vv. \vv + d \vv.d \vv $,
we have
\begin{equation}
\pint_2(\vv)
=  \pint_{1n}(\vv)
-\Bigl(\uukappa_2 \odd d\vgamma
+ d\uukappa_2.\vgamma
+ 2{D \uukappa_2 \over D t}\odd d\vv \Bigr).
\end{equation}
(For comparison purposes, the second-order Taylor development of the Cauchy tensor $ C = F^T.F $
in the neighborhood of~$ \tz $ reads
$C^\tz(\tz{+}h,\Point)
= (I+ h\,(d\vv + d\vv^T)
+ {h^2\over 2}(2d\vv^T. d\vv + d\vgamma + d\vgamma^T))(\tz,\Point) + o(h^2)$.)

\section{Visco-elasticity 2}

Second-order visco-elastic materials (not purely fluid or elastic) can be obtained with other Lie combinations.

\appendix

\def\calE{{\cal E}}

\def\Es{{E^*}}
\def\vell{{\vec\ell}}
\def\Tij{{L^i_{\;j}}}
\def\Tkm{{L^k_{\;m}}}
\def\Fs{{F^*}}
\def\Fss{{F^{**}}}
\def\tL{{\widetilde L}}
\def\Ess{{E^{**}}}
\def\calJ{{\cal J}}

\def\vB{{\vec B}}
\def\vvA{{\vv_{\!A}}}
\def\vvAt{{\vv_{\!At}}}
\def\vwA{{\vw_{\!A}}}
\def\vwAt{{\vw_{\!At}}}
\def\vvB{{\vv_{\!B}}}
\def\vvBt{{\vv_{\!Bt}}}
\def\vvBts{{\vv_{Bt*}}}
\def\calR{{\cal R}}
\def\calRA{{\calR_{\!A}}}
\def\calRB{{\calR_{\!B}}}
\def\OA{{O_{\!A}}}
\def\OB{{O_{\!B}}}
\def\ObjB{{\Obj_{\!B}}}
\def\QobjB{{Q_{Obj\!B}}}
\def\Psit{{\Psi_t}}
\def\tPhiA{{\tPhi_A}}
\def\tPsi{{\widetilde\Psi}}
\def\tPsiA{{\tPsi_{\!A}}}
\def\tPsiB{{\tPsi_{\!B}}}
\def\tPsiAt{{\tPsi_{\!At}}}
\def\tPhiB{{\tPhi_{\!B}}}
\def\tPhiAt{{\tPhi_{\!At}}}
\def\tPhiBt{{\tPhi_{\!Bt}}}
\def\tPi{{\widetilde\Pi}}
\def\pointAt{{\point_{\!At}}}
\def\pointBt{{\point_{\!Bt}}}
\def\qAt{{q_{\!At}}}
\def\qB{{q_{\!B}}}
\def\Thetat{{\Theta_t}}

\section{Tensors and objective contractions}
\label{secAtco}

Let $E$ be a finite dimensional (to simplify) real vector space, and 
let $\Es:=\calL(E;\RR)$ be the dual space of~$E$ (the space of linear functions).
An element $\ell\in \Es$ being linear, $\ell(\vv)$ is written~$\ell.\vv$, for all $\vv\in E$.
Let $\Ess:=\calL(\Es;\RR)$ be the bidual of~$E$,
and let $\calJ_1 : \vv\in E  \rar v = \calJ_1(\vv) \in \Ess$ be defined by
$v(\ell) \eqdef \ell.\vv$ for all $\ell \in E^*$.
Then $\calJ_1$ is a natural canonical isomorphism, see \eg~Spivak~\cite{SpivakACITDG},
where canonical means that the definition
only uses the constant~$1$ which is the identity element in multiplication and is the ``simplest possible''
(quite blurry),
and natural means that $\calJ_1$ is independent of the observer.
Thus $E$ and $\Ess$ are identified, and $v=\calJ_1(\vv)$ is written $v=\vv$.
And, by linearity of~$v$, $v(\ell)$ is written $v.\ell$ or~$\vv.\ell$.
(There is no natural isomorphism between $ E $ and~$ E^* $, see \eg~\cite{SpivakACITDG}:
An isomorphism exists but depends on an observer, \eg\ depends on a basis,
or \eg\ depends on an inner product see~\eref{eqr} and~\eref{eqfcv2}).

\def\Ass{A^{**}}
\def\As{A^*}
\def\Bs{B^*}
\def\Bss{B^{**}}

Let $E$ and $F$ be two finite dimensional real spaces, and $\dim E=n$ and $\dim F=m$.
Let $\calL(E;F)$ be the set of linear mappings from~$E$ to~$F$.
Let $(\va_i)$ be a basis in~$E$.
Then a linear map $L\in\calL(E;F)$ is characterized by the vectors $L.\va_j$.
Let $(\vb_i)$ be a basis in~$F$, and let $(b^i)$ be its dual basis.
And let $\Tij$ be the components of $L.\va_j$ in the basis~$(\vb_i)$, that is,
\begin{equation}
\label{eqTij1}
L.\va_j = \sumim\sumjn \Tij \vb_i,\quad\hbox{so}\quad \Tij = b^i.(L.\va_j) .
\end{equation}
And we write
\begin{equation}
\label{eqTij2}
L = \sumim\sumjn \Tij \vb_i \otimes a^j.
\end{equation}
This is coherent with the usual tensorial product:
Let $\calL(\Fs,E ; \RR)$ the space of bilinear forms on $\Fs\times E$.
If $(\vv_F,\ell_E) \in F\times \Es$, then their tensorial product $\vv_F \otimes \ell_E$
is the bilinear form in $\calL(\Fs,E ; \RR)$ defined by
$(\vv_F\otimes \ell_E)(\ell_F,\vv_E) := \vv_F(\ell_F)\,\ell_E(\vv_E) = \ell_F(\vv_F)\,\ell_E(\vv_E)$,
thanks to~$\calJ_1$, that is $(\vv_F\otimes \ell_E)(\ell_F,\vv_E) := (\ell_F.\vv_F)\,(\ell_E.\vv_E)$.
And we have the natural canonical isomorphism
$\calJ_2 : \tL \in \calL(F^*,E ; \RR) \rar L = \calJ_2(\tL) \in \calL(E; F)$
defined by, see~\cite{SpivakACITDG},
\begin{equation}
\label{eqpt2e}
\forall (\ell_F,\vu)\in F^*\times E, \quad 
\ell_F.(L.\vu) := \tL(\ell_F,\vu).
\end{equation}
Thus if $\tL = \sumim\sumjn \Tij \vb_i \otimes a^j$,
then $\Tij = \tL(b^i,\va_j)$,
then $\Tij = b^i.L.\va_j$ where $L=\calJ_2^{-1}(\tL)$,
which gives~\eref{eqTij1}.
And $[\Tij]$ is the matrix $[L]_{|\va,\vb}$ of $L\in\calL(E;F)$
as well as the matrix $[\tL]_{|\va,\vb}$ of~$\tL \in \calL(F^*,E;\RR)$ when $L=\calJ_2^{-1}(\tL)$,
relatively to the bases $(\va_i)$ and~$(\vb_i)$.
And in any case, the Einstein convention is satisfied.
In this manuscript this is applied \eg\ with
the deformation gradient $L=\Ftzt(P) = d\Phitzt(P) : \RRntz \rar \RRnt$
represented with bases in~\eref{eqFtzb}, thanks to~$\calJ_2$.

%So that $(\vb_i \otimes a^j)(b^k,\va_m) = (b^k.\vb_i)(a^j.\va_m) = \delta^k_i\delta^j_m$ and $(\sumim\sumjn \Tij \vb_i \otimes a^j)(b^k,\va_m) = \Tkm$. And we also have $b^k.(L.\va_m) = \Tkm$, see~\eref{eqTij1}.

And, with~$\calJ_2$, we have then defined the contraction
of a bilinear mapping $\tL \in \calL(F^*,E ; \RR)$
with a vector $\vu \in E$:
The definition being $\tL.\vu := L.\vu$ when $L=\calJ_2^{-1}(\tL)$.
%So~\eref{eqTij2}, where $\tL$ is meant instead of~$L$ (if we do not use~$\calJ_2$) gives~\eref{eqTij1} where $\tL$ is meant instead of~$L$ (when we use~$\calJ_2$).

Let $r, s \in \NN$ s.t. $r + s \ge 1$.
The set $\calL^r_s(E) := \calL(\underbrace{E^*,...,E^*}_{r\rm\; times}, \underbrace{E,...,E}_{s\rm\; times};\RR)$
of $ \RR $-multilinear forms
is called the set of uniform tensors of type ${r \choose s}$ on~$E$.
And $ \calL^0_0 (E) := \RR $.
An elementary uniform tensor in~$\calL^r_s(E)$ is a tensor
$ u_1 \otimes ... \otimes u_r \otimes \ell_1 \otimes ... \otimes \ell_s$,
where $ u_i \in \Ess $ and $ \ell_i \in \Es$,
which value on $(m_1,...,m_r,\vv_1,...,\vv_s)\in (\Es)^r\times E^s$
is $(u_1.m_1)...(u_r.m_r)(\ell_1.\vv_1)...(\ell_s.\vv_s)$.
And with $u_i=\calJ_1(\vu_i)$ this tensor
is written $\vu_1 \otimes ... \otimes \vu_r \otimes \ell_1 \otimes ... \otimes \ell_s$.

If $\tL_1 \in \calL^{r_1}_{s_1}(E)$ and $\tL_2 \in \calL^{r_2}_{s_2}(E)$
then their tensor product is the tensor $\tL_1 \otimes \tL_2\in \calL^{r_1+r_2}_{s_1+s_2}(E)$ defined by:
\begin{equation}
\label{eqdefpt0}
(\tL_1 \otimes \tL_2)(\ell_{1,1},...,\ell_{2,1},...,\vu_{1,1},...,\vu_{2,1},...)
: = \tL_1(\ell_{1,1},...,\vu_{1,1},...)\tL_2(\ell_{2,1},...,\vu_{2,1},...).
\end{equation}
If $\tL_1 \in \calL^{r_1}_{s_1}(E) $, $\tL_2 \in \calL^{r_2}_{s_2}(E)$,
$\ell \in \calL^0_1 (E) = \Es$, $u \in \calL^1_0(E)=\Ess$, $\vu=\calJ^{-1}(u) \in E$,
and $u$ written~$\vu$,
the objective tensor contraction of $ \tL_1 \otimes \ell \in \calL^{r_1}_{s_1+1}(E) $
with $ \vu \otimes \tL_2  \in \calL^{r_2+1}_{s_2}(E)$ is defined by
\begin{equation}
%\label{eqdefptc1}
(\tL_1\otimes \ell) . (\vu\otimes \tL_2) := (\ell . \vu)\,\tL_1 \otimes \tL_2 \in \calL^{r_1+r_2}_{s_1+s_2}(E).
\end{equation}
And the objective tensor contraction of $ \tL_1 \otimes \vu $
with $ \ell \otimes \tL_2 $ is defined by
 $(\tL_1\otimes \vu) . (\ell\otimes \tL_2) = (\vu.\ell)\,\tL_1 \otimes \tL_2
=(\ell.\vu)\,\tL_1 \otimes \tL_2$.

The objective double tensor contraction~$\odd$, for compatible tensors,
results from the simple contraction applied twice.
\Eg\ the double objective tensor contraction of
$ \tL_1 \otimes \ell_ {1,1} \otimes \vu_ {1,2} $
and $ \ell_ {2,1} \otimes \vu_ {2,2} \otimes \tL_2 $ is defined by
\begin{equation}
\label{eqdefptc2}
\eqalign{
(\tL_1\otimes \ell_{1,1}\otimes \vu_{1,2}) \odd (\ell_{2,1}\otimes \vu_{2,2}\otimes \tL_2)
= & (\vu_{1,2} . \ell_{2,1})(\ell_{1,1} . \vu_{2,2})\, \tL_1 \otimes \tL_2.
}
\end{equation}

Representation with a basis $ (\ve_i) $ of~$ E $.
\Eg\ with $S, T \in \calL^1_1(E)$,
$S = \sumjn S^i_j \ve_i \otimes e^j$ and $T = \sumjn T^i_j \ve_i \otimes e^j$,
we get
\begin{equation}
\label{eqToddS2}
S.T = \sumijkn S^i_k T^k_j \ve_i \otimes e^j, \qand
S \odd T = \sumijn S^i_j T^j_i,
\end{equation}
that is $S \odd T = \Tr(S. T)$ (the trace of~$S. T$
considered to be an endomorphism thanks to~$\calJ_2$),
which value is independent of the chosen basis (Einstein convention is satisfied).
Thus, if $\vu\in E$, $\ell\in \Es$, $T\in \calL^1_1(E)$, then $(\vu\otimes \ell) \odd T = \ell.T.\vu$;
With a basis~$(\ve_i)$, and $\vu = \sumin u^i\ve_i$, $\ell = \sumin \ell_i e^i$,
$T= \sumjn T^i_j \ve_i \otimes e^j $, we have
$(\vu\otimes \ell) \odd T
= \ell.T.\vu 
= \sumijkn \ell_i T^i_j u^j$.

\comment{
Thus if $S,T,U\in \calL^1_1(E)$, then:
\begin{equation}
\label{eqToddS2c}
(S.T)\odd U = S\odd(T.U) \quad(=\sumijkn S^i_j T^j_k U^k_i).
\end{equation}
}

Let $ A = [A^i_j] $ and $ B = [B^i_j] $ be two square $ n * n $ matrices.
The double matrix contraction of $A$ and~$B$ is defined by:
\begin{equation}
\label{eqttg}
A:B \eqdef \sumijn A^i_j B^i_j.
\end{equation}
This is not an objective contraction (the Einstein convention is not satisfied).
\Eg\ $A = [S]_{| \ve}$ and $B = [T]_{| \ve}$
give $[S]_{| \ve} : [T]_{| \ve} = A:B$, value that depends on the choice of the basis.
To be compared with $ S \odd T = A^T : B $, see~\eref{eqToddS2}.

\def\calF{{\cal F}}
Let $\calE$ be an affine space
and $E$ be the associated vector space.
Let $ \OmegaE $ be an open set in~$\calE$.
Let $ \calF(\OmegaE;\RR)$ be the set of real valued functions on~$ \OmegaE $.
A tensor of type $ {r \choose s} $ on~$ \OmegaE $, see Abraham and Marsden~\cite{AbrahamMarsden}, is a mapping
$\widetilde T : 
\point \in\OmegaE \rar \widetilde T(\point) = (\point,T(\point)) \in \OmegaE\times\calL^r_s(\OmegaE)
$
that is $\calF(\OmegaE; \RR)$-multilinear; That is,
for all $f \in \calF (\OmegaE; \RR)$
and all $z_1,z_2$ vector fields or one forms on~$\OmegaE$ where appropriate, we have
\begin{equation}
\label{eqT}
T (..., fz_1 + z_2, ...)
= f\, T (..., z_1, ...) +  T (..., z_2, ...).
\end{equation}
(That is, for all $\point\in\OmegaE$,
$T(\point) (..., f(\point)z_1(\point) + z_2(\point), ...)
= f(\point) \, T(\point) (..., z_1(\point), ...) +  T(\point) (..., z_2(\point), ...)
$.)
If there is no ambiguity then $ \widetilde T$ is simply written $T $.
The set of tensors of type ${r \choose s}$ on~$\OmegaE$ is denoted $T^r_s(\OmegaE)$.
And $T^0_0(\OmegaE) := \calF(\OmegaE; \RR)$.
Example: $T^0_1(\OmegaE)$ is identified with the set $\Omega^1(\OmegaE)$ of one-forms on~$ \OmegaE $, and
$T^1_0(\OmegaE)$ is identified with the set $\Gamma(\OmegaE)$ of vector fields on~$ \OmegaE $ thanks to~$\calJ_1$.
Counter example: A derivation operator $\nabla$ is $\RR$-linear,
but is not $\calF(\OmegaE;\RR)$-linear because \eref{eqT} is not satisfied
since $\nabla (fz_1) = (\nabla f) \, z_1 + f \, (\nabla z_1)  \ne f \, \nabla z_1$
if $f$ is not constant
($\nabla$ is not a tensor but is a spray, see~\cite{AbrahamMarsden}).

Note that the initial concept is ``vector fields'';
Then ``one-forms'' are introduced,
which are functions of vector fields;
And then ``tensors'' are introduced,
which are functions of one forms and velocity fields.

For a complete mathematical introduction (with differential geometry) see \eg\
Spivak~\cite{SpivakACITDG}, Abraham and Marsden~\cite{AbrahamMarsden},
Arnold~\cite{ArnoldODE},
Marsden and Hughes~\cite{MarsdenHughesMFE}.
%Here the simple affine case is considered.

\section{Push-forward and pull-back}
\label{secpfpb}

\def\Phittau{{\Phi^t_\tau}}
\def\ctzt{{c^\tz_{t*}}}
\def\ctz{{c_\tz}}
\def\ct{{c_t}}
\def\ctp{{c_t{}'}}
\def\ctau{{c_\tau}}
\def\cttaus{{c^t_{\tau*}}}
\def\cttaub{{c^{t*}_{\tau}}}
\def\cttaups{{(c^t_{\tau*})'}}
\def\Phit{{\Phi^t}}
\def\RRntau{{\RR^n_\tau}}

\def\vWtztb{{\vec W^{\tz*}_t}}
\def\vwt{{\vec w_t}}
\def\vwts{{\vec w_{t*}}}
\def\vwtau{{\vec w_\tau}}
\def\vwttaus{{\vec w^t_{\tau*}}}
\def\vwttaub{{\vec w^{t*}_\tau}}
\def\alphat{{\alpha_t}}
\def\alphatau{{\alpha_\tau}}
\def\alphattaus{{\alpha^t_{\tau*}}}
\def\vwttaub{{\vec w^{t*}_\tau}}
\def\alphattaub{{\alpha^{t*}_\tau}}

\def\pointtau{{\point_\tau}}
\def\pointttaus{{\point^t_{\tau*}}}
\def\pointttaub{{\point^{t*}_{\tau}}}

\def\fttaus{{f^t_{\tau *}}}
\def\fttaub{{f^{t*}_{\tau}}}
\def\PF{{P\!\!F}}
\def\PB{{P\!\!B}}

\def\Tttaub{{T^{t*}_\tau}}

\def\Omegatau{{\Omega_\tau}}
\def\Httau{{H^t_\tau}}

\def\OmegaF{{\Omega_\calF}}
\def\alphaE{{\alpha_\calE}}
\def\alphaF{{\alpha_\calF}}
\def\alphaEp{{\alpha_{\calE*}}}
\def\alphaFb{{\alpha^*_\calF}}
\def\cE{{c_\calE}}
\def\cEs{{c_{\calE*}}}
\def\cEsp{{c_{\calE*}{}'}}
\def\cF{{c_\calF}}
\def\cFb{{c^*_\calF}}
\def\fE{{f_\calE}}
\def\fEp{{f_{\calE*}}}
\def\fF{{f_\calF}}
\def\fFb{{f^*_\calF}}
\def\pE{{p_\calE}}
\def\pF{{p_\calF}}
\def\vwE{{\vw_\calE}}
\def\vwF{{\vw_\calF}}
\def\vwEp{{\vw_{\calE*}}}
\def\vwFb{{\vw^*_\calF}}

Push-forwards and pull-backs enable to define, among others,
the velocity addition formula, objectivity, and Lie derivatives.
We follow Abraham and Marsden~\cite{AbrahamMarsden}, here in a simplified affine framework.

Let $\calE$ and $\calF$ be affine sets, and let $E$ and $F$ be the associated vector space supposed to be normed,
let $\OmegaE$ and $\OmegaF$ be open sets in~$\calE$ and~$\calF$,
and let $\Psi : \OmegaE \rar \OmegaF$ be a diffeomorphism.
In this manuscript $E$ and $F$ are finite dimensional, and $\Psi$ will be either
a motion $\Phitzt : \Omegatz \rar \Omegat$, see~\eref{eqPhitzt},
or a translator $\Thetat : \calRB \rar \calRA$, see~\eref{eqdefPsit}.

The push-forward of a function $ \fE : \OmegaE \rar \RR$
by~$ \Psi $ is the function
$\Psi_*\fE = \fEp : \OmegaF \rar \RR$
defined by $\fEp:= \fE \circ (\Psi)^{-1}$, so, with $\pF = \Psi(\pE)$,
\begin{equation}
\label{eqdefpff}
\fEp(\pF) := \fE(\pE) .
\end{equation}
The pull-back of a function $\fF : \OmegaF \rar \RR$
by~$ \Psi $ is the function
$\Psi^*\fF = (\Psi^{-1})_*\fF = \fFb$, that is $\fFb = \fF \circ \Psi$, so
$\fFb(\pE) = \fF(\pF)$ when $\pF = \Psi(\pE)$.

Let 
\begin{equation}
\label{eqdefct}
\cE :
\left\{\eqalign{
]s_1,s_2[ & \rar \OmegaE \cr
s & \rar \pE=\cE(s),
}\right.
\end{equation}
be a regular curve in~$\OmegaE$.
The push-forward of~$\cE$ by~$\Psi$ is the curve
$\Psi_*\cE = \cEs$ defined by $\cEs := \Psi\circ \cE$, that is,
\begin{equation}
\label{eqcttau0}
\cEs :
\left\{\eqalign{
]s_1,s_2[ & \rar \OmegaF \cr
s & \rar \pF = \cEs(s) := \Psi(\cE(s)) \quad (=\Psi(\pE)).
}\right.
\end{equation}
Let $\cF: s\in ]s_1,s_2[ \rar \cF(s) \in \OmegaF$ be a regular curve.
The pull-back of~$\cF$ by~$\Psi$ is the curve
$\Psi^*\cF = \cFb := (\Psi^{-1})_*\cF$, that is $\cFb = \Psi^{-1}\circ \cF$. 

We use the definition of a vector field given by the tangent vectors
to a curve (see \eg~\cite{AbrahamMarsden} for equivalent definitions).
For a regular curve~$ \cE $ as in~\eref{eqdefct}, its tangent vector at $\pE=\cE(s)$ is
\begin{equation}
\vwE (\pE) \eqdef \cE' (s) .
\end{equation}
That defines the vector field $\vwE$ on~$\Im(\cE)$.
The push-forward of $\vwE$ by~$\Psi$
is the vector field $\Psi_*\vwE = \vwEp$ on~$\Im(\cEs)$
made of the tangent vectors to the push-forward curve $\cEs$, see~\eref{eqcttau0},
that is, $\vwEp(\pF) := \cEsp(s)$ when $\pF = \cEs(s)$.
So, since~\eref{eqcttau0} gives $\cEsp(s) = d\Psi(\cE(s)).\cE{}'(s)$, with $\pF = \Psi(\pE)$ we have
\begin{equation} 
\label{eqvwvW}
\vwEp(\pF) = d\Psi(\pE).\vwE(\pE) .
\end{equation}
(That is, $\vwEp = (d\Psi.\vwE)\circ\Psi^{-1}$.)
And a family of regular curves in~$\OmegaE$
gives a vector field $\vwE$ on~$\OmegaE$ and its push-forward $\vwEp$ on~$\OmegaF$.
The pull-back of a vector field $ \vwF: \OmegaF \rar F$ by~$\Psi$
is the vector field $\Psi^*\vwF = \vwFb := (\Psi^{-1})_*\vwF$, that is, with $\pF = \Psi(\pE)$,
\begin{equation}
\label{eqvwvWb}
\vwFb(\pE) := d\Psi(\pE)^{-1}.\vwF(\pF) .
\end{equation}

The push-forward of a one-form $\alphaE\in\Es$ by~$\Psi$ is the one-form $\Psi^*\alphaE = \alphaEp$
defined with $\fE = \alphaE.\vwE$, \eref{eqdefpff} and~\eref{eqvwvW};
So, with $\pF = \Psi (\pE)$,
\begin{equation}
\label{eqpalpha}
\alphaEp(\pF) := \alphaE(\pE). d\Psi(\pE)^{-1} .
\end{equation}
And the pull-back of a one-form $\alphaF \in \Fs$ by~$\Psi$ is the one-form
$\Psi^*\alphaF = (\Psi^{-1})_*\alphaF = \alphaFb$, that is, with $\pF = \Psi (\pE)$,
\begin{equation}
\label{eqpbalpha}
\alphaFb(\pE) := \alphaF (\pF). d\Psi(\pE) .
\end{equation}
This pull-back can also be defined thanks to
the natural canonical isomorphism $L \in \calL(E;F) \rar L^* \in \calL(\Fs;\Es)$
defined by $L^*(\ell_F).\vu_\calE = \ell_F.(L.\vu_\calE)$;
So $L^*(\ell_F) = \ell_F.L$;
And $L^*(\ell_F)\in \Es$ is denoted $\ell_F^*$ and named the pull-back of~$\ell_F$ by~$L$.
So that $\ell_F^* = \ell_F.L$.
And~\eref{eqpbalpha} is obtained with
$\ell_F = \alphaF(\pF)$ and $L = d\Psi(\pE)$,
and $\ell_F^*=\alphaF(\pF)^*$ is denoted $\alphaFb(\pE)$.
%(and here $E$ and $F$ are the tangent spaces at $\pE\in\OmegaE$ and at $\pF=\Psi(\pE)\in\OmegaF$).

Let $T_\calE$ be a ${r \choose s}$ tensor on~$\OmegaE$, with $r,s\ge 1$.
Its push-forward by $\Psi$ is the ${r \choose s}$ tensor $\Psi_*T_\calE=T_{_\calE*}$ on~$\OmegaF$ defined by
\begin{equation}
T_{_\calE*}(\alpha_{1\calF},...,\alpha_{r\calF},\vw_{1\calF},...,\vw_{s\calF})
= T_\calE(\alpha_{1\calF}^*,...,\alpha_{r\calF}^*,\vw_{1\calF}^*,...,\vw_{s\calF}^*).
\end{equation}
Let $T_\calF$ be a ${r \choose s}$ tensor on~$\OmegaF$, with $r,s\ge 1$.
Its pull-back by $\Psi$ is the ${r \choose s}$ tensor $T_\calF^*$ on~$\OmegaE$ defined by
\begin{equation}
T_\calF^*(\alpha_{1\calE},...,\alpha_{r\calE},\vw_{1\calE},...,\vw_{s\calE})
= T_\calF(\alpha_{1\calE*},...,\alpha_{r\calE*},\vw_{1\calE*},...,\vw_{s\calE*}).
\end{equation}

\section{Lie derivatives: Introduction and interpretation}

\label{secdlfei}

\def\ctpush{{c_{t*}}}
\def\ctaupull{{c^*_\tau}}

(Classical mechanics.)
Let $\tPhi$ be a motion, see~\eref{eqtPhi}, $\vv$ be the associated Eulerian velocity, see~\eref{eqdefvv},
$t,\tau\in\RR$, $\Phittau$ the associated motion, see~\eref{eqPhitzt0} and~\eref{eqPhitzt},
and $\Fttau = d\Phittau$.
With reference to \S~\ref{secpfpb},
let $\Psi = \Phittau$, $\OmegaE = \Omegat$, $\OmegaF = \Omegatau$, $\pE = \pointt$ and $\pF=\pointtau = \Phittau(\pointt)$.
If $f_\tau : \Omegatau \rar \RR$ is a function then
its pull-back $(\Phittau)^*f_\tau = \fttaub$ is given by
$\fttaub(\pointt) = f_\tau(\pointtau)$.
If $\vwtau$ is a vector field in~$\Omegatau$ then
its pull-back $(\Phittau)^*\vwtau = \vwttaub$ is given by 
$\vwttaub(\pointt) = \Fttau(\pointt)^{-1}.\vwtau(\pointtau)$.
If $\alphatau$ is one-form in~$\Omegatau$ then
its pull-back $(\Phittau)^*\alphatau = \alphattaub$ is given by 
$\alphattaub(\pointt) = \alphatau(\pointtau). \Fttau(\pointt)$.
And we define $f^*_\tau(t,\pointt):= \fttaub(\pointt)$,
$\vw^*_\tau(t,\pointt) := \vwttaub(\pointt)$,
and $\alpha^*_\tau(t,\pointt) := \alphattaub(\pointt)$.

The Lie derivative of a real-valued function along the flow of velocity~$\vv$
(or along the motion~$\tPhi$) at $(t,\pointt)$ is defined by
\begin{equation}
\label{eqdefLief2}
\calL_\vv f(t,\pointt) \eqdef \lim_{\tau\rar t} {f^*_\tau(t,\pointt) - f(t,\pointt) \over \tau-t}.
\end{equation}
Thus, with $\pointtau = \Phittau(\pointt)$,
\begin{equation}
\label{eqdefLief2b}
\calL_\vv f(t,\pointt) = \lim_{\tau\rar t} {f(\tau,\pointtau) - f(t,\pointt) \over \tau-t}
= {Df\over Dt}(t,\pointt).
\end{equation}
{\bf Interpretation:}
An observer does not have the gift of spatial and/or temporal ubiquity, thus cannot trivially compare values
at a two distinct points or instants: He needs time (from $t$ to~$\tau$)
and a displacement (from $\pointt$ to~$\pointtau$) to compare $f(\tau,\pointtau)$ and $f(t,\pointt)$
(to compute the difference $f(\tau,\pointtau) - f(t,\pointt)$).
So~\eref{eqdefLief2b} is a computation consequence of~\eref{eqdefLief2}.

The Lie derivative $\calL_\vv\vw$ of a vector field $\vw$ along the flow of velocity~$\vv$ is defined by
\begin{equation}
\label{eqdefLiev2}
\calL_\vv\vw(t,\pointt)
\eqdef \lim_{\tau\rar t} {\vw^*_\tau(t,\pointt) - \vw(t,\pointt) \over \tau-t},
\end{equation}
and the difference $\vw^{t*}(t,\pointt) - \vw(t,\pointt)$
is computed at a single time~$t$ and at a single point~$\pointt$ (does not require ubiquity).
In $\RRn$, $\calL_\vv\vw$ is equivalently defined by
\begin{equation}
\label{eqdefLiev}
\calL_\vv\vw(t,\pointt) \eqdef \lim_{\tau\rar t} {\vw(\tau,\pointtau)-\vw^t_*(\tau,\pointtau) \over \tau-t}
%\quad\hbox{or}\quad =  \lim_{h\rar 0} {\vw_t(\pointt) - ((\Phi^{t{-}h}_t)_*\vw_{t-h})(\pointt) \over h}
.
\end{equation}
{\bf Interpretation:}
\Eg\ in~$ \RRn $ with~\eref{eqdefLiev}:
at~$\tau$ and $\pointtau$, the numerator $\vw(\tau,\pointtau) - \vw^t_*(\tau,\pointtau)$
compares the true value $\vw(\tau,\pointtau)$ of~$\vw$ at $(\tau, \pointtau)$ with
the value $\vw^t_*(\tau,\pointtau)$ which corresponds to
``the vector that would have let itself transported by the motion'' (the push-forward).
So the Lie derivative $\calL_\vv\vw(t,\pointt)$,
which is the limit of the rate ${\vw(\tau,\pointtau)-\vw^t_*(\tau,\pointtau) \over \tau-t}$,
measure the ``resistance'' of~$ \vw $ ``submitted to the motion~$\tPhi$''
(``submitted to the flow'').
The Lie Virtual Power Principle proposed in this manuscript is based on this property.
(For comparison with the classical approach see~\S~\ref{remspqsv}.)

And~\eref{eqdefLiev2} gives:
\begin{equation}
\label{eqdefLiev3}
\calL_\vv \vw = {D\vw \over Dt} - d\vv.\vw \quad ( = {\pa \vw\over \pa t} + d\vw.\vv - d\vv.\vw).
\end{equation}
Indeed with $\pointtau = \Phittau(\pointt)$ and
$\vg: \tau \rar \vg(\tau) = d \Phittau (\pointt)^{-1}.\vw(\tau,\pointtau)$
we have $ \calL_ \vv \vw(t,\pointt) = \vg \, '(t) $, see~\eref{eqdefLiev2};
And $ d \Phi^t (\tau, \pointt). \vg (\tau) = \vw (\tau, \pointtau) $ leads to
$ d {\pa \Phi^t \over \pa \tau} (\tau, \pointt). \vg (\tau) + d \Phi^t (\tau, \pointt). \vg \, '( \tau)
= {D \vw \over D \tau} (\tau, \pointtau) $,
therefore with $ \tau = t $ we get
$ d \vv (t, \pointt). \vw (t, \pointt) + \vg \, '(t) = {D \vw \over Dt} (t, \pointt) $,
thus~\eref{eqdefLiev3}.
Note that, for a vector field~$\vw$ submitted to the flow, the term $ - d\vv.\vw$ in $\calL_\vv \vw$
takes into account the influence of non uniform flows~$\vv$ (flows s.t. $d\vv\ne0$) on~$\vw$.

The Lie derivative $\calL_\vv \alpha$ of a one-form $\alpha$ along the flow of velocity~$\vv$ is the one-form 
deduced for example from~\eref{eqdefLiev3} and \eref{eqdefLief2b} with
the derivation property $ \calL_ \vv (\alpha. \vw) = \calL_ \vv \alpha. \vw + \alpha. \calL_ \vv \vw $. Thus
\begin{equation}
\label{eqpfalpha}
\calL_ \vv \alpha = {D \alpha \over Dt} + \alpha.d \vv \quad (= {\pa \alpha \over \pa t} + d \alpha. \vv + \alpha.d \vv).
\end{equation}
The same result is obtained with the mathematical definition
$ \calL_ \vv \alpha (t, \pointt)
:= \lim_{\tau \rar t} {\alpha^*_\tau(t,\pointt) - \alpha(t,\pointt) \over \tau-t}$.

And for tensors of order~$ 2 $,  $ \uukappam \in \Tuuot $ (mixed), $ \uukappau \in \Tdzot $ (up)
and $ \uukappad \in \Tzdot $ (down), the generic derivation property
$ \calL_ \vv (a \otimes b) = (\calL_ \vv a) \otimes b + a \otimes \calL_ \vv b$
(or definition with pull-backs) gives
\begin{equation}
\label{eqdl}
\eqalign{
&\calL_\vv\uukappam
= {D\uukappam \over D t}
+ \uukappam.d\vv - d\vv.\uukappam
\; \in \Tuuot
\quad \hbox{(Jaumann)}, 
\cr
&\calL_\vv\uukappau
= {D \uukappau \over D t}
- \uukappau.(d\vv)^* - d\vv.\uukappau 
\; \in \Tdzot
\quad \hbox{(Maxwell upper-convected)}, 
\cr
&\calL_\vv\uukappad
= {D \uukappad \over D t}
+ \uukappad.d\vv + (d\vv)^*.\uukappad
\; \in \Tzdot
\quad \hbox{(Maxwell lower-convected)}, 
\cr
}
\end{equation}
where, for $\uusigmam \in \Tuuot$, the adjoint tensor $ {\uusigmam}^* \in \Tuuot^*$
is defined by $ {\uusigmam}^* (\vu, \ell) = \uusigmam (\ell, \vu)$,
and ${D \uukappa \over D t} = {\pa \uukappad \over \pa t} + d\uukappa.\vv$ (generic notation).
%(Similar definitions for tensors of any order.)

Although an Eulerian velocity field $ \vv $ is not objective, see~\S~\ref{secpf} and~\ref{secco},
we have:
If $T$ is an objective tensor,
then its Lie derivative $ \calL_ \vv T$ is an objective tensor;
see \eg\ Marsden and Hughes~\cite{MarsdenHughesMFE} p.101.
(But partial derivatives and material derivative are not objective in general.)
Here objectivity refers to ``covariant objectivity'', see~\eref{eqwobj}.

\section{Lie derivative versus deformation tensor}
\label{remspqsv}

The deformation tensor $ C = F^T.F $
between $\tz$ and~$t$ is used to measure the relative deformation between two~vectors
thanks to the use of two Euclidean inner products, $\dd_G$ at~$\tz$ and $\dd_g$ at~$t$
(see \eg\ Marsden and Hughes~\cite{MarsdenHughesMFE}):
With $\Point \in \Omegatz$, $\pointt = \Phitzt(\Point)$,
$\Ftzt(\Point) := d \Phitzt(\Point)$,
$\Ctzt(\Point) = (\Ftzt(\Point))^T. \Ftzt(\Point)$,
$\vW_i(\Point) \in \RRntz$ and
$\vw_{i*}(\pointt) = \Ftzt(\Point). \vW_i(\Point)$
the push-forward of~$\vW_i$ by~$ \Phitzt $ see~\eref{eqvwvW},
we have :
\begin{equation}
\label{eqspqsv}
\eqalign{
(\vw_{1*}(\pointt),\vw_{2*}(\pointt))_g
= & (\Ctzt(\Point).\vW_1(\Point),\vW_2(\Point))_G.
}
\end{equation}
This value is compared with $ (\vW_1 (\Point), \vW_2 (\Point))_G $ in classical mechanics.
N.B.: The deformation tensor~$C$ compares two vectors that have let themselves
deformed by the flow, see~\eref{eqspqsv}, since $\vw_{1*}$ and $\vw_{2*}$ are the push-forwards
by the flow.

While the Lie derivative of a vector field $\vw$
measures the resistance of a single vector field $\vw$ submitted to a flow,
see interpretation of~\eref{eqdefLiev},
and does not require a priori the use of inner products (Euclidean or not)
since there is no comparison between two vectors.

\section{Change of Riesz representation vector}
\label{secthmriesz}

The Riesz representation theorem is often implicitly used in classical mechanics under
isometric objectivity hypotheses (see~\S~\ref{secco}). Unfortunately this causes problems
since covariance cannot be confused with contravariance.
\Eg, Misner, Thorne, Wheeler~\cite{MisnerThorneWheeler} box~2.1:
``Without it [the distinction between covariance and contravariance],
one cannot know whether a vector is meant or the very different geometric object that is a 1-form.''
%(One forms are defined on vector fields.)

Let $E$ be a normed vector space and $E^*$ the set of linear continuous functions $E\rar \RR$.
There is no natural canonical isomorphism between $E$ and its dual~$\Es$,
see \eg~Spivak~\cite{SpivakACITDG},
but if an observer introduces an inner product then an isomorphism (depending on the observer) is obtained:

\begin{theorem}[Riesz representation theorem]
\label{thmriesz}
If $\dd_g$ is an inner product in~$E$ so that $E$ is a Hilbert space, then
\begin{equation}
\label{eqr}
\forall\ell\in \Es,\quad\exists ! \vell_g \in E,\quad \forall \vw \in E,\quad \ell.\vw = (\vell_g,\vw)_g.
\end{equation}
Moreover $||\vell_g||_g = ||\ell||_\Es$.
And the vector $\vell_g$ is called
the~Riesz representative vector of the linear form~$\ell $ relatively to the inner product~$\dd_g$,
or the $\dd_g$-Riesz representative vector of~$\ell$.
\end{theorem}

Proof: If $\ell=0$ then $\vell_g=0$.
If $\ell\ne0$, choose a
$\vv\notin {\rm Ker}\ell=\{\vv\in E : \ell(\vv)=0\} = \ell^{-1}(\{0\})$
(closed sub-space since $\ell$ is continuous). 
Let $\vv_0$ be the $\dd_g$-orthogonal projection of~$\vv$ on~${\rm Ker}\ell$,
and let $\vn = {\vv-\vv_0 \over ||\vv-\vv_0||_g}$ (a $||.||_g$-unitary vector normal to~${\rm Ker}\ell$).
Then take $\vell_g = \ell(\vn)\vn$ to get~\eref{eqr} and
$||\vell_g||_g = \sup_{||\vw||_E=1}|\ell.\vw|$ ($||\ell||_\Es$) thanks to the Cauchy--Schwarz inequality.
And uniqueness is trivial. %~\qed

\begin{corollary}[Change of Riesz representation vector]
\label{propcv}
Let $ \ell \in \Es $.
Let $ \dd_a $ and $ \dd_b $ be two inner products,
and let $\vell_a$ and $\vell_b$ be the $\dd_a$ and $\dd_b$-Riesz representative of~$\ell$. Then
\begin{equation}
\label{eqfcv0}
\forall \vw\in E,\quad (\vell_a,\vw)_a = (\vell_b,\vw)_b .
\end{equation}
\end{corollary}

Proof: For all~$\vw$, \eref{eqr} gives
$\ell.\vw = (\vell_a,\vw)_a$ and $\ell.\vw = (\vell_b,\vw)_b $, thus~\eref{eqfcv0}.

\begin{example}
\label{exafcv2}
Let $ \dd_a $ be the Euclidean inner product built with the English foot~(ft)
and $ \dd_b $ be the Euclidean inner product built with the meter~(m).
We have $1$~ft $= \mu$~m with $\mu= 0.3048$.
And $\dd_b = \mu^2 \, \dd_a$ (and $||.||_b = \mu||.||_a$),
thus~\eref{eqfcv0} gives $ (\vell_a, \vw) _a = \mu^2 (\vell_b, \vw) _a $ for all~$ \vw $,
therefore:
\begin{equation}
\label{eqfcv2}
\vell_a = \mu^2\vell_b
\end{equation}
and the vector $ \vell_a $ is $ \mu^2 $ times smaller
(more than ten times smaller) than the vector~$\vell_b$.
%(Another proof of~\eref{eqfcv2} is given with the $\dd_g$-unit normal vectors in the proof of theorem~\ref{thmriesz}.)
So the Riesz representation vector depends on the choice of the inner product
(an inner product is a measuring tool, and a change of tool changes the result).
%Thus a Riesz representation vector cannot be used if (covariant) objectivity is under concern.
%(Unfortunately, a Riesz representation vector is overused in classical mechanics although a linear form cannot be identified with a vector.)
\end{example}

\section{Incompatibilities with Riesz representation vectors}
\label{secincomp}

\def\valpha{{\vec\alpha}}
\def\vbetag{{\vec\beta_g}}
\def\alphap{{\alpha_*}}
\def\valphapg{{\valpha_{*g}}}

We have just seen that the Riesz representation vector depends on the observer, see e.g.~\eref{eqfcv2}.
But we also have e.g.:

1- Incompatibility with push-forwards.
Let $\alpha$ be in $(\RRntz)^*$ and
let $\alphap = \alpha.\Ftzt^{-1} \in (\RRnt)^*$ be its push-forward by~$\Phitzt$, see~\eref{eqpalpha}.
Let $\dd_G$ and $\dd_g$ be inner products in~$\RRntz$ and~$\RRnt$.
Let $\valpha_G\in\RRntz$ and $\valphapg\in\RRnt$ be the
$\dd_G$ and $\dd_g$-Riesz representation vectors of $\alpha$ and $\alphap$, see~\eref{eqr}.
And let $\valpha_{G*} \in\RRnt$ be the push-forward of $\valpha_G$ by~$\Phitzt$, see~\eref{eqvwvW}.
So, with $p=\Phitzt(P)$ and $\vw_p\in\RRnt$, we have $(\valphapg(p),\vw_p)_g = \alphap(p).\vw_p
= (\alpha(P).\Ftzt(P)^{-1}).\vw_p
= \alpha(P).(\Ftzt(P)^{-1}.\vw_p)	
= (\valpha_G(P),\Ftzt(P)^{-1}.\vw_p)_G
= (\Ftzt(P)^{-T}.\valpha_G(P),\vw_p)_g
$, true for all~$\vw_p$, thus
\begin{equation}
\valphapg(p) = \Ftzt(P)^{-T}.\valpha_G(p).
\end{equation}
So $\valphapg$ is not the push-forward of~$\valpha_G$,
that is $\valphapg(p) \ne \Ftzt(P).\valpha_G(p)$, unless $\Phitzt$ is the motion of a solid.
Thus the Riesz representation vectors should not be used if push-forwards of one-forms are needed.

2- Incompatibility with Lie derivative.
Let $\beta$ be in $(\RRnt)^*$.
Let $\dd_g$ be an inner product in~$\RRnt$,
and let $\vbetag$ be the $\dd_g$-Riesz representation vector of~$\beta$,
that is $\beta.\vw = (\vbetag,\vw)_g$ for any~$\vw$.
Then we have $\calL_\vv\beta.\vw \ne (\calL_\vv \vbetag,\vw)_g$, unless $\Phi$ is the motion of a solid,
that is we have ${D\beta \over Dt}.\vw + \beta.d\vv.\vw
\ne ({D\vbetag \over Dt},\vw)_g - (\vbetag,d\vv.\vw)_g$ in general.
We have:
\begin{equation}
\calL_\vv \beta.\vw = (\calL_\vv \vbetag,\vw)_g + (\vbetag,(d\vv+d\vv^T).\vw)_g
\end{equation}
since ${D\beta \over Dt}.\vw + \beta.d\vv.\vw
= ({D\vbetag\over Dt},\vw)_g + (\vbetag,d\vv.\vw)_g$.
Thus the Riesz representation vectors
cannot be used if Lie derivative of one-forms are needed.

\section{Velocity addition formula}
\label{secpf}

(This \S\ is needed to define objectivity, see the next~\S.)
Classical mechanics setting.
The observers use the same time scale.
An observer $A$ defines a referential $ \calRA = (\OA, (\vA_i)) $
and an observer $B$ defines a referential $ \calRB = (\OB, (\vB_i))$,
with $(\vA_i)$ and $(\vB_i)$ Cartesian bases.
They observe an object~$\Obj$ in a time interval $[t_1,t_2]$.
Let $\tPhiA: [\tu,\td] \times \Obj  \rar \calRA$
and $\tPhiB:[\tu,\td] \times \Obj \rar \calRB$ be the motions of~$\Obj$ as described by $A$ and $B$,
and let $\vv_A$ and $\vv_B$ be the associated Eulerian velocity fields, see~\eref{eqtPhi} and~\eref{eqdefvv},
that is,
\begin{equation}
\label{eqphiAB}
\left\{\eqalign{
& \pointAt = \tPhiA(t,\Pobj) = \OA+\sumin x_{At}^i \vec A_i, \cr
& \vv_A(t,\pointAt) = {\pa \tPhiA \over \pa t}(t,\Pobj),
}\right.
\quad
\left\{\eqalign{
& \pointBt = \tPhiB(t,\Pobj) = \OB+\sumin x_{Bt}^i \vec B_i,\cr
& \vv_B(t,\pointBt) = {\pa \tPhiB \over \pa t}(t,\Pobj).
}\right.
\end{equation}
Let $\ObjB$ be the object used by~$B$ to define his referential
(\eg\ $\ObjB =$ Earth).
Let $\tPsiA: [\tu,\td] \times \ObjB  \rar \calRA$
and $\tPsiB:\ObjB \rar \calRB$ (static) be the motions of~$\ObjB$ described by $A$ and $B$,
and let $\vw_A$ and $\vw_B$ be the associated Eulerian velocity fields, see~\eref{eqtPhi} and~\eref{eqdefvv},
that is,
\begin{equation}
\label{eqtPsiA}
\left\{\eqalign{
& \qAt = \tPsiA(t,\QobjB)= \OA+\sumin y_{At}^i \vec A_i, \cr
& \vw_A(t,\qAt) = {\pa \tPsiA \over \pa t}(t,\QobjB),
}\right.
\quad
\left\{\eqalign{
& \qB = \tPsiB(\QobjB) = \OB+\sumin y_B^i \vec B_i, \cr
& \vw_B(t,\qB)=\vec0.
}\right.
\end{equation}
For $t$ fixed, let $\tPhiAt(\Pobj) := \tPhiA (t,\Pobj)$ and $\tPhiBt(\Pobj) := \tPhiB (t,\Pobj)$,
and $\tPsiAt(\Pobj) := \tPsiA (t,\Pobj)$.

The mapping $\tPhiA$, $\tPsiA$ in~$\calRA$ and $\tPhiB$, $\tPsiB$ in~$\calRB$
are motions: They are defined by one observer in his referential.
The translation mapping at~$t$ from $B$ to~$A$ (the translator) is the ``inter-referential'' diffeomorphism
$\Thetat:=\tPsiAt\circ\tPsiB^{-1}$, and we denote
\begin{equation}
\label{eqdefPsit}
\Thetat :
\left\{\eqalign{
\calRB & \rar  \calRA \cr
\qB = \OB+\sumin y_B^i \vec B_i & \rar \qAt = \Thetat(\qB) = \OA+\sumin y_{At}^i \vec A_i
}\right\}  .
\end{equation}
So, $\tPsiAt(\QobjB) = \Thetat(\tPsiB(\QobjB))$ for all $\QobjB\in\ObjB$:
If a particle $\QobjB$ of $\ObjB$ is located at $\qB = \tPsiB(\QobjB)\in\calRB$ by the observer~$B$,
then the observer~$A$ locates $\QobjB$ at~$t$ at $\tPsiAt(\QobjB) = \qAt = \Thetat(\qB)\in\calRA$.
\Eg\ $\Thetat(\OB)$ is ``the position of $\OB$ in~$\calRA$ at~$t$'';
And thus the push-forward $d\Thetat(\OB).\vB_i$ of $\vB_i$ by~$\Thetat$ is
``the basis vector $\vB_i$ as seen by~$A$ at~$\Thetat(\OB)$''
(see~\S~\ref{secpfpb} with the diffeomorphism $\Psi = \Thetat$
and $\cE=$ the $i$-th coordinate line in~$\calRB$).

With~\eref{eqdefPsit} define $ \Theta: [\tu, \td] \times \calRB \rar \calRA $
by $ \Theta (t, \qB) \eqdef \Thetat (\qB)$.
And define the ``$\Theta$-Eulerian velocity'' at $\qAt \in\calRA$ by:
\begin{equation}
\label{eqvvrat}
\vw_\Theta (t,\qAt) = {\pa \Theta \over \pa t} (t, \qB) \qif \qAt = \Thetat(\qB).
\end{equation}
Note that $\vw_\Theta$ looks like a Eulerian velocity but is not
since $\Theta$ is not a motion ($\Thetat$~is an inter-referential mapping).

\comment{
So, for all $\Pobj\in\Obj$ and $\QobjB\in\ObjB$, we have
\begin{equation}
\label{eqloidec}
\left\{\eqalign{
&(\qAt =) \quad \tPsiA(t,\QobjB) = \Theta(t,\tPsiB(\QobjB)) \quad (=\Theta(t,\qB)), \cr 
&(\pointAt =) \quad \tPhiA(t,\Pobj) = \Theta(t,\tPhiB(t,\Pobj)) \quad (=\Theta(t,\pointBt)). \cr 
}\right.
\end{equation}
}

{\bf Interpretation of~$\vw_\Theta$:}
With $\tPsiA(t,\QobjB) = \Theta(t,\tPsiB(\QobjB))$ and $\qB = \tPsiB(\QobjB)$ we get
${\pa\tPsiA \over \pa t}(t,\QobjB) = {\pa\Theta \over \pa t}(t,\qB)$,
thus, with \eref{eqtPsiA} and~\eref{eqvvrat} %and $\qAt = \Thetat(\qB)$, 
we have
\begin{equation}
\label{eqiwt}
\vw_A(t,\qAt) = \vw_\Theta(t,\qAt),
\end{equation}
equality in~$\calRA$; And the ``$\Theta$-Eulerian velocity'' $\vw_\Theta(t,\qAt)$
is the Eulerian velocity $\vw_A(t,\qAt)$ of the particle $\QobjB\in\ObjB$ that
is at~$t$ at $\qAt = \tPsiAt(\QobjB) = \Thetat(\qB) \in \calRA$.

{\bf Velocity addition formula.}
We also have $\tPhiAt = \Thetat\circ\tPhiBt$, that is, with~\eref{eqphiAB}, $\pointAt = \Thetat(\pointBt)$
(inter-referential relations between positions at~$t$).
So $\tPhiA(t,\Pobj) = \Theta(t,\tPhiB(t,\Pobj))$, and with $\pointBt = \tPhiB(t,\Pobj)$ we get
${\pa\tPhiA \over \pa t}(t,\Pobj)
= {\pa\Theta\over \pa t}(t,\pointBt) + d\Theta(t,\pointBt).{\pa\tPhiB\over \pa t}(t,\Pobj)$,
that is, $\vv_A(t,\pointAt) = \vw_\Theta(t,\pointAt) + d\Theta(t,\pointBt).\vv_B(t,\pointBt)$,
that is with~\eref{eqiwt},
\begin{equation}
\label{eqloicv2}
\vvAt(\pointAt) = \vvBts(\pointAt) + \vwAt(\pointAt) 
\qwhere \vvBts (\pointAt) \eqdef d \Thetat (\pointBt). \vvBt (\pointBt).
\end{equation}
This formula is the velocity addition formula.
Interpretation: At~$t$, if $A$ is the ``absolute observer'' and 
$B$ is the ``relative observer'' then the formula reads:
``$(\vvAt$ the absolute velocity) = ($\vvBts$ the relative velocity translated for~$A$)
+ ($\vwAt$ the velocity of the %Cartesian 
coordinate system of~$B$ in~$\calRA$)'', equality in~$\calRA$.

\section{Covariant and isometric objectivities}
\label{secco}

\def\vuA{{\vu_{\!A}}}
\def\vuB{{\vu_B}}
\def\vuAt{{\vu_{\!At}}}
\def\vuBt{{\vu_{\!Bt}}}
\def\vuBts{{\vu_{\!Bt*}}}

(Mainly from Marsden and Hughes~\cite{MarsdenHughesMFE}.) Setting of the previous~\S.

\begin{definition}
Let $\vuBt$ be a vector field in~$\calRB$ at~$t$ as described by the observer~$B$.
Let $\vuBts$ be its translation for~$A$,
that is, the push-forward of~$\vuBt$ by~$\Thetat$
(so $\vuBts(\pointAt) = d\Thetat(\pointBt).\vuBt(\pointBt)$ when $\pointAt = \Thetat(\pointBt)$, see~\eref{eqvwvW}).
Then $\vuBts$ is called the objective transform of $\vuBt$ by~$\Thetat$.
More generally, at~$t$ the objective transform of a tensor $T_{Bt}$ in~$\calRB$
is its push-forward $T_{Bt*}$ by~$\Thetat$.
\end{definition}

\begin{definition}
\label{defov}
Consider ``a quantity'' that can be described as a vector field by any observer,
quantity written~$\vu$.
At~$t$, $A$~describes $\vu$ as the vector field $\vuAt$ in~$\calRA$,
and $B$~describes $\vu$ as the vector field $\vuBt$ in~$\calRB$.
Then $\vu$ is (covariant) objective if and only if,
for all $t$ and all observers $A$ and~$B$,
$\vuAt$ is the objective transform of $\vuBt$ by~$\Thetat$,
that is,
\begin{equation}
\label{eqwobj}
\vuAt = \vuBts.
\end{equation}
In other words, $\vu$ is objective iff,
for all $t$ and all observers $A$ and~$B$, whenever $\qAt = \Thetat(\qB)$ we have
$\vuAt(\qAt) = d\Thetat(\qB).\vuBt(\qB)$.
Similarly, a tensor $T$ is objective iff $T_{At} = T_{Bt*}$ for all $t$ and all observers.
\end{definition}

Counter-example:
a velocity field $\vv$ is not objective, see~\eref{eqloicv2}: $\vwAt \ne0$ in general.
So the objective vector fields in the definition~\ref{defov} refer, \eg, to the ``force fields'' $\vf$
%that act on the material and 
of Newton's fundamental law $\sum \vf = m\vgamma$ where $\vgamma = {D\vv \over Dt}$ (acceleration).

\begin{definition}
To work in an objective covariant framework means to consider any diffeomorphism~$\Thetat$ (any translator).
\end{definition}

\begin{definition}
To work in an ``objective isometric'' framework means to only consider
observers $A$ and $B$ using a unique metric~$\dd_g$ (isometric setting),
that is, to only consider the $\Thetat$ that are isometries relatively to one chosen
inner product~$\dd_g$
(and $(\Thetat.\vu_1,\Thetat.\vu_2)_g = (\vu_1,\vu_2)_g$ for any $\vu_1,\vu_2$).
(And in classical mechanics the chosen metric~$\dd_g$ is often assumed to a Euclidean metric.)
\end{definition}

\Eg\ an English observer using feet and a french observer using meters
have no choice but to use covariant objectivity if they want to communicate (they use different metrics).
And in general relativity ``objective isometry'' is meaningless.
%(Isometric objectivity is the starting point of the ``frame invariance principle'', see~\S~\ref{secintro} and Truesdell et Noll~\cite{TruesdellNoll2004}.)

\section{About elasticity and classical formulations}
\label{secsel}

The appendix follows Germain~\cite{GermainX1984}, apart from the remarks where $\Tr(F)$
is replaced by $\det(F)$.

$\bullet$ In $\RRntz$.
The abbreviated notations
$ C = F^T.F $ (deformation tensor)
and $ E = {C-I \over 2} $ (Green-Lagrange tensor)
stand for the endomorphisms in~$\RRntz$ relative to~$\tz$, $t$ and $\Point\in\Omegatz$ defined by
$ \Ctzt (\Point) = \Ftzt (\Point)^T. \Ftzt (\Point) $
and $ \Etzt (\Point) = {\Ctzt (\Point) -I_ \tz \over 2} $ where $I_\tz$ is the identity in~$\RRntz$.
The classical isotropic homogeneous elasticity is \eg\ stated as,
\begin{equation}
\label{equs1}
\uusigma = \lambda \Tr(E) \,I + 2 \mu E,
\end{equation}
%where $\lambda, \mu \in \RR$,
abbreviated notation for the endomorphism
$ \uusigmatzt (\Point) = \lambda \Tr (\Etzt (\Point)) \, I + 2 \mu \Etzt (\Point) $ in~$\RRntz$.
%Since a Eulerian velocity field $\vv$ and its differential $d \vv$ refer to~$\Omegat$, and $\uusigma$ refers to $\Omegatz$, an objective product as $\uusigma \odd d \vv $, see~\eref{eqh1}, cannot be directly considered.

Small deformations:
The linearization $ E = {C-I \over 2} \simeq {F + F^T \over 2} -I $ gives (linear elasticity)
\begin{equation}
\label{equs1p}
\uusigma \simeq \lambda \Tr(\uueps) \,I + 2 \mu \uueps, \quad \uueps = {F + F^T \over 2}-I.
\end{equation}
But $ \uueps $ has no functional meaning
because $ \Ftzt (P): \RRntz \rar \RRnt $, $ \Ftzt (P)^T: \RRnt \rar \RRntz $,
and $ I=I_\tz $ stays in $\RRntz$, so that the sum $ {\Ftzt + \Ftzt^T \over 2} -I_\tz $ is not a function
(and $ \Tr (\eps) $ has no functional meaning either),
unless the shifter is introduced, see remark~\ref{remshifter0},
and/or \eref{equs1p} is considered in the matrix sense after having chosen
a unique Euclidean basis at~$\tz$ and~$t$.

\def\Jtzt{{J^\tz_t}}
\def\Jtz{{J^\tz}}

Remark relative to the introduction of the virtual power of pressure~\eref{eqcalPpres}:
Relatively to Euclidean bases,
the volume change at $\Point\in\Omegatz$ 
is $J = \det (F) = (\det(C))^\demi$,
where $J$, $F$ and $C$ stands for $\Jtzt(\Point)$, $\Ftzt(\Point)$ and $\Ctzt(\Point)$.
%And, $\vv$ being the Eulerian velocity, the rate of volumic dilatation at $\pointt = \Phitzt(\Point)$ is $\dvg\vv(t,\pointt) = {1\over \Jtz(t,\Point)}{\pa\Jtz(t,\Point) \over \pa t}$.
And for small displacements $\Tr (C-I) \simeq \det (C) -1$:
Indeed with an orthonormal basis of diagonalization of~$C$
we have $ [C] = \diag (1 {+} \eps_1, ..., 1 {+} \eps_n)$,
and then $ \Tr (C) = n + \sum_i \eps_i $ and $ \det (C) = 1+ \sum_i \eps_i + o (\eps)$
in the neighborhood of $ \eps = 0 $ where $ \eps = \max (|\eps_i|)$.
So $ \Tr (E) = {\Tr (C) -n \over 2} \simeq {\det (C) - 1 \over 2}$.
Then, for small displacements, \eref{equs1} can be replaced by
\begin{equation}
\label{equs1p2}
\uusigma =  {\lambda\over 2} (\det(C)-1) \,I +  \mu (C-I).
\end{equation}
%where $ \lambda, \mu \in \RR $.
And $\det(C) = \det(F^T.F) = \det(F)^2$, and
$\det (F)^2 -1 = (\det(F)-I)(\det(F)+I) \simeq 2(\det(F)-I) $ for small displacements.
Then a pressure term can be considered with~\eref{eqcalPpres}.

$\bullet$ In $ \RRnt $.
Instead of the deformation tensor $C = F^T.F$ we may prefer to use the Finger tensor $ \uub = F.F^T$,
and more precisely its inverse
$\uub^{- 1} = H^T.H $ where $ H = F^{- 1}$.
Unabbreviated notation: At $t$ and $\pointt = \Phitzt(\Point)$,
$\Htzt (\pointt) = \Ftzt (\Point)^{- 1}$ and
$(\uubtzt)^{- 1} (\pointt) = \Htzt(\pointt)^T . \Htzt (\pointt)$.
And instead of the Green--Lagrange tensor, we may use the Euler--Almansi tensor defined by
\begin{equation}
\label{equs20}
\uua
= {I_t - \uub^{-1} \over 2}
= {I_t - H^T.H \over 2},
\end{equation}
simplified notation of the endomorphism
$ \uuatzt (\pointt) = {I_t - (\uubtzt)^{- 1} (\pointt) \over 2} $ in~$ \RRnt $.
The classical elasticity is then \eg\ stated as, with $ \lambda, \mu \in \RR $,
\begin{equation}
\label{equs2}
\uusigma 
= \lambda \Tr(\uua) \,I + 2 \mu \uua,
\end{equation}
to compare with~\eref{equs1}.
And here $ \uusigma \odd d \vv $ is meaningful (double objective contraction in~$\RRnt$).
For small displacements, and for matrix computation, \eref{equs20} yields
$ [\uua] 
\simeq [I] - {[H] + [H]^T \over 2} $ (linearization),
and~\eref{equs2} reads (linear elasticity)
\begin{equation}
\label{equs2p}
[\uusigma] = \lambda \Tr([\uua]) \,I + 2 \mu [\uua] \qavec [\uua] = [I]-{[H]+[H]^T \over 2}].
\end{equation}
Since $ [\Ftz (t, \Point)] - [I] = [I] - [\Htz (t, \point (t))] + o (t {-} \tz) $ if $ \point (t) = \Phitz (t, \Point) $,
we get back to~\eref{equs1p}.
And, as for~\eref{equs1p2}, instead of~\eref{equs2} we can consider, for small displacements,
\begin{equation}
\label{equs2pb}
\uusigma = {\lambda \over 2} (1 - \det(\uub^{-1})) \,I + \mu (I - \uub^{-1}).
\end{equation}
(Or with $(1 - \det (H))$ instead of~$\Tr (I-H)$.)
And a pressure term can be considered with~\eref{eqcalPpres}.

\bibliography{biblio}

\end{document}